\documentclass[final,3p,times]{elsarticle}

\usepackage{amsmath,amssymb,amsfonts}
\usepackage{graphicx}
\graphicspath{{figures/}}
\usepackage{booktabs}
\usepackage[hidelinks]{hyperref}
\usepackage{url}
\usepackage{subcaption}
\usepackage{bm}
\usepackage{siunitx}
\usepackage{float}
\usepackage[section]{placeins}
\usepackage{rotating}
\usepackage{pdflscape}

\begin{document}

\begin{frontmatter}

\begin{highlights}
\item Proposed a physics-guided neural operator surrogate for discretization-invariant RANS prediction around arbitrary airfoils.
\item Reduced velocity field MSE by up to 23× compared to baseline neural operator surrogates.
\item Introduced Learned Canonical Quadrature, reducing drag prediction error by 7.5×.
\item Achieved approximately 10,000× speedup over conventional RANS solvers, enabling rapid surrogate-based aerodynamic design and optimization.
\end{highlights}

\title{DD-RNO: A Domain-Decomposed Routed Neural Operator for Airfoil Flow Prediction}
\author[ref1]{T. A. Mehta}
\author[ref1]{P. S. Bhati}
\author[ref2]{H. D. Akolekar\corref{cor1}}
\address[ref1]{Department of Computer Science \& Engineering, Indian Institute of Technology Jodhpur, Jodhpur, India}
\address[ref2]{Department of Aerospace Engineering, Indian Institute of Technology Jodhpur, Jodhpur, India}
\cortext[cor1]{Corresponding author}

\begin{abstract}
Deep learning surrogates for RANS flow prediction around airfoils face two persistent bottlenecks. A single neural architecture cannot simultaneously resolve sharp near-wall boundary layers and smooth far-field potential flow. Additionally, force prediction is undermined by the numerical instability of computing wall-normal velocity gradients from continuous-field approximations. 
 Both of these points are addressed with a DD-RNO (domain-decomposed routed neural operator), combining a spectral geometry encoder with two physics-guided innovations: (a) a differentiable domain routing mechanism that partitions the flow field into inviscid, boundary-layer, and wake regimes
---dispatching query points to specialized regional decoders, and (b) learned canonical quadrature (LCQ), which replaces unstable pressure integration with flow-conditioned, learned integration weights that predict lift and drag directly from surface pressure. 
On the AirfRANS benchmark, DD-RNO cuts velocity field mean-square error (MSE) by 17$\times$ ($u_x$) and 12$\times$ ($u_y$) over the strongest baseline, widening to 23$\times$ under out-of-distribution Reynolds extrapolation---evidence that the routing mechanism generalizes with the physics it encodes rather than merely fitting the training distribution. LCQ reduces drag MSE by 7.5$\times$ relative to conventional pressure integration and raises drag rank correlation from $\rho = 0.250$ to $\rho = 0.997$. Ablations confirm that both components are indispensable to performance: removing domain routing increases velocity error by 8.2$\times$, and removing LCQ increases relative drag error more than 40-$\times$. At $\sim$144\, ms per sample---a 10{,}000$\times$ speedup over conventional RANS solvers---DD-RNO offers a surrogate accurate and fast enough for real-time aerodynamic design and optimization loops.
\end{abstract}


\end{frontmatter}

\section*{Nomenclature}
\addcontentsline{toc}{section}{Nomenclature}

\begin{tabular}{@{}llp{6.5cm}@{}}
\multicolumn{3}{@{}l}{\textbf{Roman Symbols}} \\
$c$ & Airfoil chord length & [\si{\meter}] \\
$C_D$ & Total drag coefficient & [--] \\
$C_L$ & Total lift coefficient & [--] \\
$C_{D,0}$ & Zero-lift drag coefficient & [--] \\
$C_p$ & Surface pressure coefficient & [--] \\
$m/c$ & Airfoil maximum camber ratio & [\%] \\
$p$ & Pressure field & [\si{\pascal}] \\
$\text{Re}$ & Reynolds number based on chord length & [--] \\
$t/c$ & Airfoil maximum thickness ratio & [\%] \\
$u_x, u_y$ & Streamwise and transverse velocity components & [\si{\meter\per\second}] \\
$U_\infty$ & Free-stream velocity magnitude & [\si{\meter\per\second}] \\
$\mathbf{w}$ & Global flow latent conditioning state & [--] \\
$\mathbf{W}_{\text{canon}}$ & Learned canonical quadrature integration weights & [--] \\
$\mathbf{x} = (x, y)$ & Query coordinate vector in physical domain & [\si{\meter}] \\
\\[-2pt]
\multicolumn{3}{@{}l}{\textbf{Greek Symbols}} \\
$\alpha$ & Angle of attack & [\si{\degree}] \\
$\delta_{\text{BL}}$ & Local turbulent boundary layer thickness envelope & [\si{\meter}] \\
$\nu_t$ & Turbulent kinematic eddy viscosity & [\si{\meter\squared\per\second}] \\
$\Phi(\mathbf{x})$ & Signed Distance Function (SDF) to airfoil surface & [\si{\meter}] \\
$\chi_{\text{BL}}, \chi_{\text{wake}}, \chi_{\text{inv}}$ & Spatial domain decomposition routing weights & [--] \\
\\[-2pt]
\multicolumn{3}{@{}l}{\textbf{Acronyms}} \\
BL & Boundary Layer & \\
CFD & Computational fluid dynamics & \\
CNN & Convolutional neural network & \\
DD-RNO & Domain-decomposed routed neural operator & \\
FiLM & Feature-wise linear modulation & \\
FNO & Fourier neural operator &  \\
GELU & Gaussian error linear unit & \\
GNN & Graph neural network & \\
LCQ & Learned canonical quadrature & \\
LGI & Learned geometric integration & \\
MAE & Mean Absolute Error & \\
MLP & Multilayer perceptron & \\
MSE & Mean squared error & \\
OOD & Out-of-distribution & \\
PDE & Partial differential equation & \\
PINN & Physics-informed neural network & \\
POD & Proper orthogonal decomposition & \\
RANS & Reynolds-averaged Navier-Stokes & \\
SDF & Signed distance function & \\
\end{tabular}

\section{Introduction}
\label{sec:introduction}

Computational fluid dynamics (CFD) simulations are the cornerstone of industrial aerodynamic analysis. However, aerodynamic design often requires thousands of CFD evaluations, resulting in high computational costs and extended design cycles. Consequently, even industry-standard CFD approaches can become prohibitively expensive for iterative tasks such as shape optimization, uncertainty quantification, and design space exploration.
This has motivated substantial interest in data-driven surrogate models that can approximate CFD solutions at a fraction of the cost \cite{ribeiro2020deepcfd, brunton2020mlfluidmechanics}.

The aerospace community has a long-standing tradition of utilizing surrogate models to replace expensive CFD simulations within design optimization loops \cite{forrester2009surrogate, queipo2005surrogate}. Early approaches relied on response surface methods, Kriging, and radial basis functions to approximate scalar quantities of interest directly from geometric parameters. While highly effective for low-dimensional design spaces, these classical methods were incapable of reconstructing full flow fields. To extend surrogate modeling to full-field reconstruction, projection-based reduced-order models built on POD were developed \cite{braconnier2011pod, mifsud2010pod}. POD-based surrogates represent the flow field as a linear combination of dominant modes extracted from an ensemble of CFD data. Although these methods achieve significant computational efficiency, their linear and global formulations struggle to capture strongly nonlinear phenomena like flow separation or shock waves.

The advent of deep learning introduced convolutional architectures to RANS flow-field prediction. Bhatnagar et al. \cite{bhatnagar2019cnn} trained a convolutional neural network (CNN) on rasterized signed-distance representations of airfoil geometries to predict velocity and pressure fields, demonstrating massive speedups over traditional CFD. Thuerey et al. \cite{thuerey2020deepflow} employed modern U-Net architectures to study the coupled effects of training dataset size and network capacity. Similarly, Sekar and Khoo \cite{sekar2019fastflow} achieved comparable predictive accuracy using a CNN trained on a similar rasterized geometry representation. Grid-based CNNs established the viability of learning full RANS fields straight from geometry, but because their outputs rely on a fixed resolution, they are bound to the specific pixel grid used to train them, and must be re-trained for new grid resolutions or sizes.

A parallel line of research focuses on data-driven turbulence modeling, where machine learning is used to augment or replace classical turbulence closures \cite{akolekar2019machine, yilmaz2021turbulence, Akolekar2021,pacciani2021assessment,Fang2026}. These strategies learn corrections to the Reynolds stress tensor or eddy viscosity from high-fidelity data and are typically coupled back into an iterative solver. Turbulence-related quantities exhibit a structured, data-supported regularity that can be systematically exploited to enhance predictive accuracy.

Graph neural networks (GNNs) have become a standard framework for physics simulations on unstructured meshes. By aggregating information from neighboring nodes, GraphSAGE \cite{hamilton2017graphsage} enabled models to generalize to entirely new graph structures. To capture multi-scale features across irregular domains, Graph U-Nets \cite{gao2019graphunet} introduced hierarchical graph pooling. MeshGraphNets \cite{pfaff2021meshgraphnets} then applied this approach directly to physical systems, showing that learned graph solvers can accurately mimic complex dynamics on unstructured meshes. Building on this idea, Sanchez-Gonzalez et al. \cite{sanchez2020gns} demonstrated that graph-based models could simulate a wide variety of physics, ranging from fluids to rigid bodies. Yet, despite these strengths, mesh-based GNNs are still tied to their underlying graph topology. Because predictions only happen at fixed node locations, the message-passing mechanism naturally incorporates assumptions about local connectivity and mesh resolution. As a result, these models cannot be queried at arbitrary, continuous coordinates. This is a major drawback for tasks that rely on adaptive sampling or resolution-independent predictions.The AirfRANS benchmark \cite{bonnet2023airfrans} established a standardized evaluation protocol for such methods and provided the baseline performance metrics against which subsequent developments are evaluated.

Operator learning approaches \cite{kovachki2021neuraloperator} brought a new approach to surrogate modeling by learning direct mappings between infinite-dimensional function spaces, unlike neural networks which process finite dimensional vectors. This helps them break free from the discretization dependencies of traditional neural networks. For example, the Fourier Neural Operator (FNO) \cite{li2021fno} parameterizes integral kernel operators in the Fourier domain, enabling discretization-invariant predictions and smooth generalization across different grid resolutions when solving PDE systems. On the other hand, DeepONet \cite{lu2021deeponet} pairs a branch network with a trunk network to map input functions straight to output functions. To keep these models grounded in real physics, physics-informed variants \cite{goswami2022pino} integrate PDE residual losses into training. To tackle irregular shapes, Geo-FNO \cite{li2022geofno} maps physical domains onto a uniform computational grid via a learned geometric deformation, making fast Fourier transforms (FFTs) usable on non-rectangular geometries. However, relying on these deformation maps introduces approximation errors and often struggles around sharp geometric features, such as the leading edge of an airfoil. On top of that, spectral methods naturally struggle with sharp, highly localized details like boundary layers \cite{rahaman2019spectral}—which is precisely why using specialized decoders for different regions of a flow field is so attractive.

Coordinate-based neural fields, also known as implicit neural representations (INRs), have gained significant traction in computer vision and graphics for representing shapes, radiance fields, and complex scenes. These networks parameterize functions as continuous mappings from spatial coordinates to output values, achieving complete resolution independence by construction.
SIREN \cite{sitzmann2020siren} demonstrated that periodic activation functions allow coordinate networks to accurately represent high-frequency signal details that standard Multi-Layer Perceptrons (MLPs) with ReLU activations typically fail to capture due to spectral bias. This insight was further extended by Tancik et al. \cite{tancik2020fourier}, who showed that mapping input coordinates to high-dimensional sinusoidal features prior to the standard MLP layers yielded similar performance benefits. Consequently, Fourier feature networks have become a standard tool for learning functions with high-frequency variations.

For physical systems, Physics-Informed Neural Networks (PINNs) \cite{raissi2019pinn} use coordinate-based architectures to solve PDEs directly, incorporating the governing physical equations straight into the loss function. To help standard PINNs tackle larger, more complex geometries, Extended PINNs \cite{jagtap2020xpinn} introduced space-time domain decomposition, assigning dedicated subnetworks to different subdomains. Even so, both PINNs and their domain-decomposed variants consistently struggle with enforcing tricky boundary conditions and scaling up to turbulent regimes \cite{sun2020physics}. Scaled frameworks like SimNet \cite{hennigh2021simnet} and DeepXDE \cite{lu2021deepxde} have made training these models much easier to work with, but problems with high-Reynolds-number flows persist. When it comes to aerodynamic surrogate modeling, CORAL \cite{serrano2023coral} introduced an encode-process-decode framework that compresses geometric representations—such as SDF grids and decodes them directly into physical fields at specific query coordinates. More recently, MARIO \cite{catalani2025mario} streamlined this process by cutting out the separate latent-processing step, delivering strong results on the AirfRANS data-scarce benchmark.

Despite these advances, standard coordinate-based networks have a fundamental problem; a single network with uniform capacity is forced to compromise between over-smoothing sharp near-wall gradients and introducing unwanted high-frequency noise in the far field. A domain-decomposed architecture can solve this exact trade-off. By routing queries to specialized decoders, the network matches its representation capacity directly to the distinct spectral demands of each flow region.

To summarize, current surrogate modeling approaches come with clear trade-offs: i)mesh-based GNNs deliver high accuracy, but they remain fundamentally tied to the input grid discretization, ii)FNOs provide discretization invariance, but they depend heavily on structured computational grids, iii)standard coordinate-based networks offer excellent geometric flexibility, yet they struggle to resolve multi-scale physical features simultaneously.

A separate challenge is integrating aerodynamic forces. Lift and drag are obtained by integrating surface pressure and shear stress, quantities sensitive to boundary-layer resolution. Multi-task optimization with volumetric and force supervision often leads to trade-offs where one task dominates learning. Furthermore, conventional pressure integration with fixed geometric weights is sensitive to discretization, while viscous drag requires computing $\partial u / \partial n|_{\text{wall}}$, which is numerically unstable when derived from predicted velocity fields \cite{arzani2021wss, tikhonov1977illposed}.

The DD-RNO framework is designed to overcome these limitations through three core innovations. This framework builds upon and extends a rich body of literature as discussed above.  The proposed innovations represent a novel methodology within the context of aerodynamic surrogate modeling to the best of the authors' knowledge.

First, a \textbf{hybrid spectral-implicit architecture} pairs the global contextual awareness of spectral operators with the local querying flexibility of coordinate-based neural fields. An FNO-based spectral geometry encoder processes the continuous geometric input to build a spatial feature map. Individual query points are then enriched by combining multi-scale Fourier positional encodings with features locally probed from this map using bilinear interpolation. This design effectively balances global geometric context with sharp local resolution across arbitrary, discretization-independent coordinates.

Second, a \textbf{physics-guided domain routing} mechanism splits the flow field into three physically distinct regions: the inviscid outer flow, the boundary layer, and the trailing wake. Rather than forcing a single network to resolve all physical scales simultaneously, smooth, differentiable gates driven by wall distance and turbulent boundary layer scaling ($\delta_{\text{BL}} \propto \text{Re}^{-1/5}$) route query points to specialized regional decoders. This structural partitioning allows each decoder to specialize in its targeted flow region.

Third, \textbf{learned canonical quadrature (LCQ)} is introduced as a novel force-prediction module within the DD-RNO framework. Traditional surrogates typically estimate forces either through direct scalar regression (which discards spatial integration context) or standard numerical pressure integration (which remains sensitive to grid discretization and relies on unstable wall-normal velocity gradients). LCQ resolves this dilemma by mapping predicted surface pressure distributions onto a standardized canonical manifold while learning flow-conditioned integration weights. This mechanism enables the model to integrate surface pressure distributions directly into total lift ($C_L$) and drag ($C_D$) coefficients while implicitly accounting for geometry-dependent viscous drag.

DD-RNO is designed to address these core challenges on two-dimensional aerodynamic flow benchmarks \cite{bonnet2023airfrans} by establishing a unified, discretization-invariant framework for simultaneous spatial field prediction and continuous force integration.
The remainder of this paper is organized as follows. Section~\ref{sec:methodology} presents the DD-RNO architecture in detail. Section~\ref{sec:results} describes the experimental setup and presents quantitative and qualitative results.  Section~\ref{sec:conclusion} concludes the paper and discusses future directions.

\section{Methodology}
\label{sec:methodology}

\begin{sidewaysfigure*}[p]
    \centering
    \includegraphics[width=1.1\textwidth]{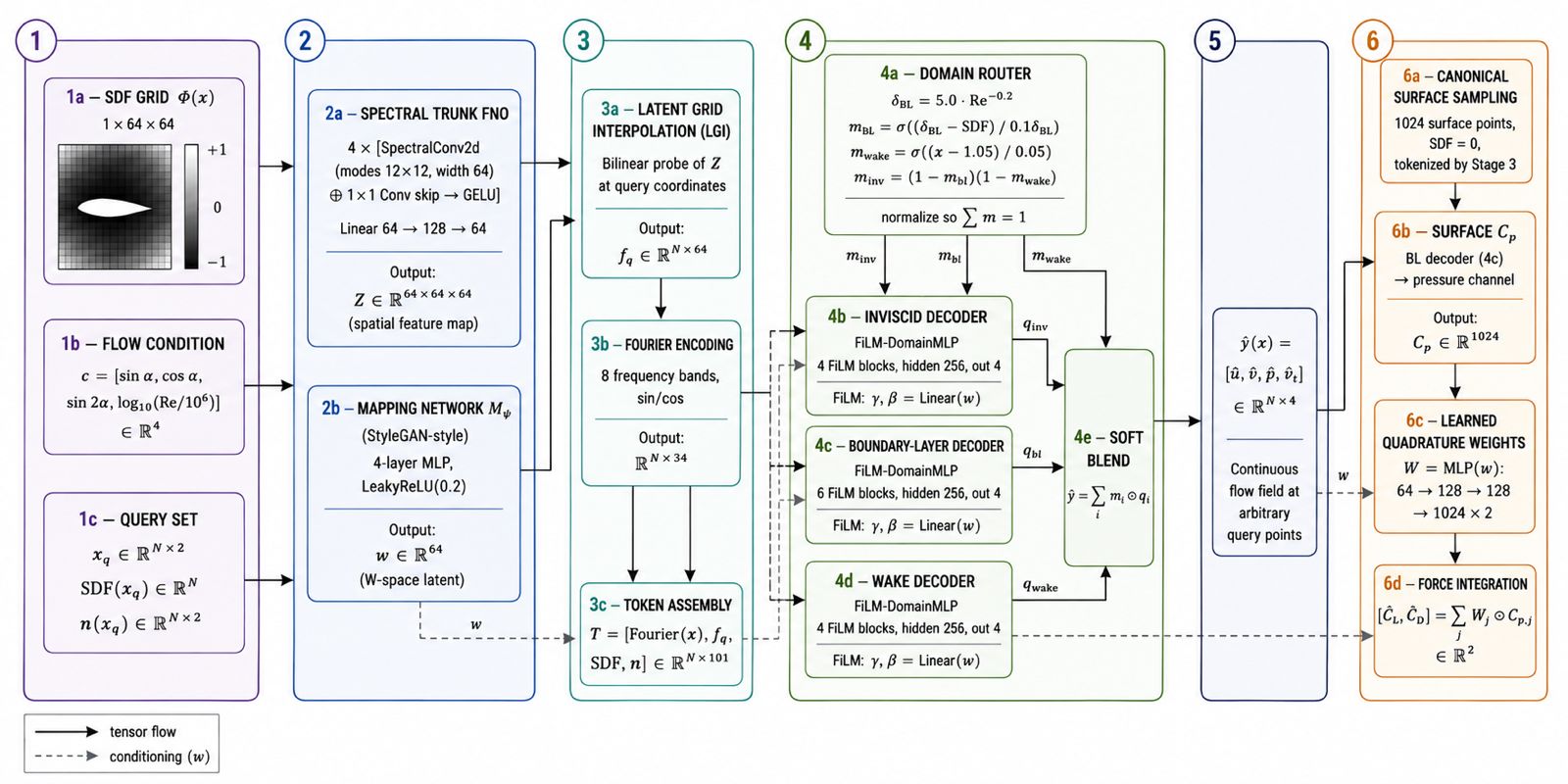}
    \caption{Overview of the DD-RNO architecture.}
    \label{fig:architecture}
\end{sidewaysfigure*}

Figure~\ref{fig:architecture} provides a high-level overview of the DD-RNO pipeline. Each component is described in detail below, explaining the physical and engineering decisions behind design choices.

\subsection{\textbf{Dataset}}
\label{sec:dataset_envelope}

The proposed DD-RNO architecture is trained and evaluated on the AirfRANS dataset \cite{bonnet2023airfrans}. The dataset consists of 1,000 steady-state 2D RANS simulations computed using unstructured OpenFOAM meshes with over $200{,}000$ nodes, utilizing the Spalart-Allmaras turbulence model for closure. The operational parameter space spans:
\begin{itemize}
    \item {Reynolds number ($\mathrm{Re}$) range}: $\mathrm{Re} \in [2\times10^6, 6\times10^6]$,
    \item {Angle of attack ($\alpha$) range}: $\alpha \in [-5^\circ, 15^\circ]$.
\end{itemize}

\subsection{\textbf{Problem Formulation}}
\label{sec:problem}

Let $\Omega \subset \mathbb{R}^2$ denote the fluid domain bounded internally by the solid airfoil surface $\partial\Omega_{\text{wall}}$ and externally by the far-field boundary $\partial\Omega_{\infty}$. The steady, incompressible RANS equations are:
\begin{align}
\nabla\cdot\mathbf{u} &= 0, \label{eq:continuity} \\
(\mathbf{u}\cdot\nabla)\mathbf{u} + \frac{1}{\rho}\nabla p - \nabla\cdot\left[\nu_{\mathrm{eff}}(\nabla\mathbf{u}+\nabla\mathbf{u}^{\top})\right] &= \mathbf{0}, \label{eq:momentum} \\
\nu_{\mathrm{eff}}(\mathbf{x}) &= \nu+\nu_t(\mathbf{x}), \label{eq:nu_eff}
\end{align}
where Eq.~\eqref{eq:continuity} is the continuity equation, Eq.~\eqref{eq:momentum} is the steady-state momentum equation, and Eq.~\eqref{eq:nu_eff} defines the effective kinematic viscosity as the sum of molecular kinematic viscosity $\nu$ and turbulent eddy viscosity $\nu_t$.

The Spalart-Allmaras (SA) model \cite{spalart1992sa} is selected for turbulence closure because it is specifically optimized for external aerodynamic applications and is the standard model used to generate the ground-truth simulations in the AirfRANS dataset. Under steady-state conditions, the model solves a transport equation for a working turbulent variable $\tilde{\nu}$:
\begin{equation}
(\mathbf{u} \cdot \nabla) \tilde{\nu} = c_{b1} \tilde{S} \tilde{\nu} + \frac{1}{\sigma} \left[ \nabla \cdot \left((\nu + \tilde{\nu})\nabla \tilde{\nu}\right) + c_{b2} |\nabla \tilde{\nu}|^2 \right] - c_{w1} f_w \left(\frac{\tilde{\nu}}{s}\right)^2, \label{eq:sa_transport}
\end{equation}
where $s$ (abbreviated from $s(\mathbf{x}, \partial\Omega_{\text{wall}})$) is the shortest distance to the closest solid wall boundary $\partial\Omega_{\text{wall}}$, and the turbulent eddy viscosity is computed as $\nu_t = \tilde{\nu} f_{v1}$ with $f_{v1} = \chi^3 / (\chi^3 + c_{v1}^3)$ and $\chi = \tilde{\nu}/\nu$.

The physical boundary conditions are formulated as follows:
\begin{equation}
\mathbf{u}|_{\partial\Omega_{\text{wall}}} = \mathbf{0}, \quad \mathbf{u}|_{\partial\Omega_{\infty}} = U_\infty(\cos\alpha, \sin\alpha)^\top,
\end{equation}
where $\mathbf{u}|_{\partial\Omega_{\text{wall}}} = \mathbf{0}$ represents the physical no-slip and no-penetration boundary conditions on the solid wall (implying both normal and tangential velocity components vanish at the wall), and $U_\infty = 1$ is the non-dimensionalized (normalized) freestream velocity at the far-field boundary $\partial\Omega_{\infty}$ with the angle of attack $\alpha$ within the dataset range.

The learning objective is to approximate the mapping:
\begin{equation}
\mathcal{F}_\theta: \left( \Phi(\mathbf{x}), \alpha, \mathrm{Re} \right) \mapsto \left( \hat{u}_x(\mathbf{x}), \hat{u}_y(\mathbf{x}), \hat{p}(\mathbf{x}), \hat{\nu}_t(\mathbf{x}) \right),
\end{equation}
where $\Phi: \mathbb{R}^2 \to \mathbb{R}$ is the SDF representing the airfoil geometry, defined by $|\nabla\Phi| = 1$, $\Phi < 0$ inside the airfoil, and $\Phi = 0$ on the wall.

\subsection{\textbf{Geometry and Flow Encoding}}
\label{sec:encoding}

\subsubsection{Signed Distance Function}

To represent the arbitrary shape of an airfoil in a format suitable for deep learning, the SDF \cite{park2019deepsdf}, denoted by $\Phi(\mathbf{x})$, is utilized. Conceptually, the SDF is a scalar field where the value at any spatial coordinate $\mathbf{x}$ represents the shortest Euclidean distance to the airfoil boundary $\partial\Omega_{\text{wall}}$:
\begin{equation}
\Phi(\mathbf{x}) = \begin{cases}
-s(\mathbf{x}, \partial\Omega_{\text{wall}}) & \text{if } \mathbf{x} \text{ is inside the airfoil}, \\
0 & \text{if } \mathbf{x} \in \partial\Omega_{\text{wall}}, \\
s(\mathbf{x}, \partial\Omega_{\text{wall}}) & \text{if } \mathbf{x} \text{ is outside the airfoil},
\end{cases}
\end{equation}
where $s(\mathbf{x}, \partial\Omega_{\text{wall}}) = \inf_{\mathbf{y}\in\partial\Omega_{\text{wall}}} \|\mathbf{x} - \mathbf{y}\|$ represents the shortest Euclidean distance from the query point $\mathbf{x}$ to the solid wall boundary $\partial\Omega_{\text{wall}}$. Here, the infimum $\inf$ denotes the greatest lower bound of the distance set, representing the absolute minimum distance from the coordinate point $\mathbf{x}$ to any point $\mathbf{y}$ on the boundary curve $\partial\Omega_{\text{wall}}$.

SDF is chosen as the primary geometric representation instead of a binary occupancy grid or a list of boundary coordinates for three reasons:
\begin{enumerate}
    \item{Continuous spatial context}: Unlike binary grids, which introduce sharp discontinuities at boundaries that degrade neural network convergence, the SDF provides a smooth, continuous gradient field. This ensures that every point in the flow domain is immediately aware of its relative proximity and orientation to the solid wall.
    \item{Physical alignment with aerodynamics}: Fluid flow physics, particularly boundary layer growth and viscous shear stresses, are fundamentally governed by the distance to the solid wall. By explicitly embedding this distance metric ($\Phi(\mathbf{x})$) into the input space, the network can naturally learn region-specific scaling laws without needing to compute them from coordinates.
    \item{Analytical geometric features}: Because the SDF is continuous and differentiable, geometric properties like the wall-normal unit vector $\hat{\mathbf{n}}(\mathbf{x})$ can be computed analytically at any coordinate:
    \begin{equation}
    \hat{\mathbf{n}}(\mathbf{x}) = \frac{\nabla\Phi(\mathbf{x})}{\|\nabla\Phi(\mathbf{x})\|}.
    \end{equation}
    This eliminates the need for geometric discretization or mesh-based approximations when resolving surface pressures and forces.
\end{enumerate}

For numerical processing, the SDF is discretized on a $64 \times 64$ Cartesian grid spanning the chord-normalized domain $[-0.5, 1.5] \times [-1, 1]$. This resolution was selected through a grid convergence study: dropping to a coarser $32 \times 32$ grid degraded FNO feature quality (boosting velocity MSE by 42\%), whereas stepping up to $128 \times 128$ offered only marginal gains (a 3\% MSE reduction) at four times the computational cost.

The spectral geometry encoder processes this $64 \times 64$ grid primarily to extract global, low-frequency geometric features like airfoil camber and thickness. The high-frequency, near-wall details and steep boundary-layer gradients are instead handled by continuous, coordinate-based decoders. These decoders query continuous coordinates $\mathbf{x}$ directly, receiving the exact continuous distance $\Phi(\mathbf{x})$ and analytical normal vectors $\hat{\mathbf{n}}(\mathbf{x})$. Paired with multi-scale Fourier features that act as high-pass filters, this setup enables the decoders to capture sharp velocity gradients at sub-grid precision—effectively decoupling the resolution of the geometry encoder from the physical scale of the boundary layer.

\subsubsection{Flow Condition Encoding}

To stabilize training, the input physical parameters are mapped into a standardized numerical range. First, the angle of attack $\alpha$ is encoded using trigonometric functions:
\begin{equation}
\mathbf{e}_\alpha = (\sin\alpha, \cos\alpha, \sin 2\alpha)^\top.
\end{equation}
Conditioning neural networks on raw angles often introduces numerical instabilities near periodic boundaries. Mapping $\alpha$ to sine and cosine terms ($\sin\alpha, \cos\alpha$) projects the angle onto a continuous circular coordinate system, preserving continuity across the domain. The third term, $\sin 2\alpha$, is explicitly included to capture second-harmonic physical trends. Aerodynamic forces and integrated coefficients are strongly tied to quadratic trigonometric relationships such as lift scaling with $\sin 2\alpha$ across broad operating ranges, or induced drag scaling with $\sin^2\alpha = \frac{1-\cos 2\alpha}{2}$. Feeding the double-frequency harmonic $\sin 2\alpha$ directly into the network lets it map these second-order geometric projections and non-linear transitions without needing to approximate the trigonometric identity internally.

Second, because Reynolds numbers in external aerodynamics span several orders of magnitude ($2\times 10^6$ to $6\times 10^6$), the Reynolds number is log-compressed to linearize its range:
\begin{equation}
e_{\mathrm{Re}} = \log_{10}(\mathrm{Re} / 10^6).
\end{equation}
This logarithmic scaling provides the network with uniform sensitivity to flow transitions across both lower and higher Reynolds number regimes. The final flow condition vector is:
\begin{equation}
\mathbf{c} = [\mathbf{e}_\alpha; e_{\mathrm{Re}}] \in \mathbb{R}^4.
\end{equation}

\subsubsection{Multi-Scale Fourier Features}

When neural networks are trained to map continuous spatial coordinates $\mathbf{x} = [x, y]^\top$ to physical fields, they suffer from a well-documented phenomenon known as "spectral bias" \cite{rahaman2019spectral}. As a result of this, they naturally prioritize learning smooth, low-frequency patterns, making it difficult to resolve sharp, high-frequency details like thin boundary layers near the airfoil surface.

To overcome this bottleneck, query coordinates are mapped into a high-dimensional space using multi-scale Fourier features \cite{tancik2020fourier}:
\begin{equation}
\mathbf{\gamma}(\mathbf{x};\mathbf{B}) = [\sin(\mathbf{B}^\top\mathbf{x}); \cos(\mathbf{B}^\top\mathbf{x})],
\end{equation}
where $\mathbf{B}$ is a projection matrix containing a log-spaced spectrum of spatial frequencies spanning eight scales ($L=8$). For a 2D spatial coordinate $\mathbf{x} = [x, y]^\top \in \mathbb{R}^2$, applying sine and cosine projections across these eight frequency scales produces $8 \times 2 \times 2 = 32$ features. Concatenating these spectral terms with the 2 raw spatial coordinates yields a 34-dimensional positional embedding $\mathbf{\gamma}_{\text{MS}}(\mathbf{x}) \in \mathbb{R}^{34}$. By projecting coordinates across multiple scales simultaneously, this mapping effectively acts as a multi-band spectral filter. This allows the downstream decoders to resolve both the slowly varying potential flow field (low frequencies) and the sharp near-wall velocity gradients (high frequencies) simultaneously. This embedding is passed directly as input to each regional decoder.

\subsection{\textbf{Spectral Geometry Encoder}}
\label{sec:sge}

\begin{figure}[!ht]
    \centering
    \includegraphics[width=\columnwidth]{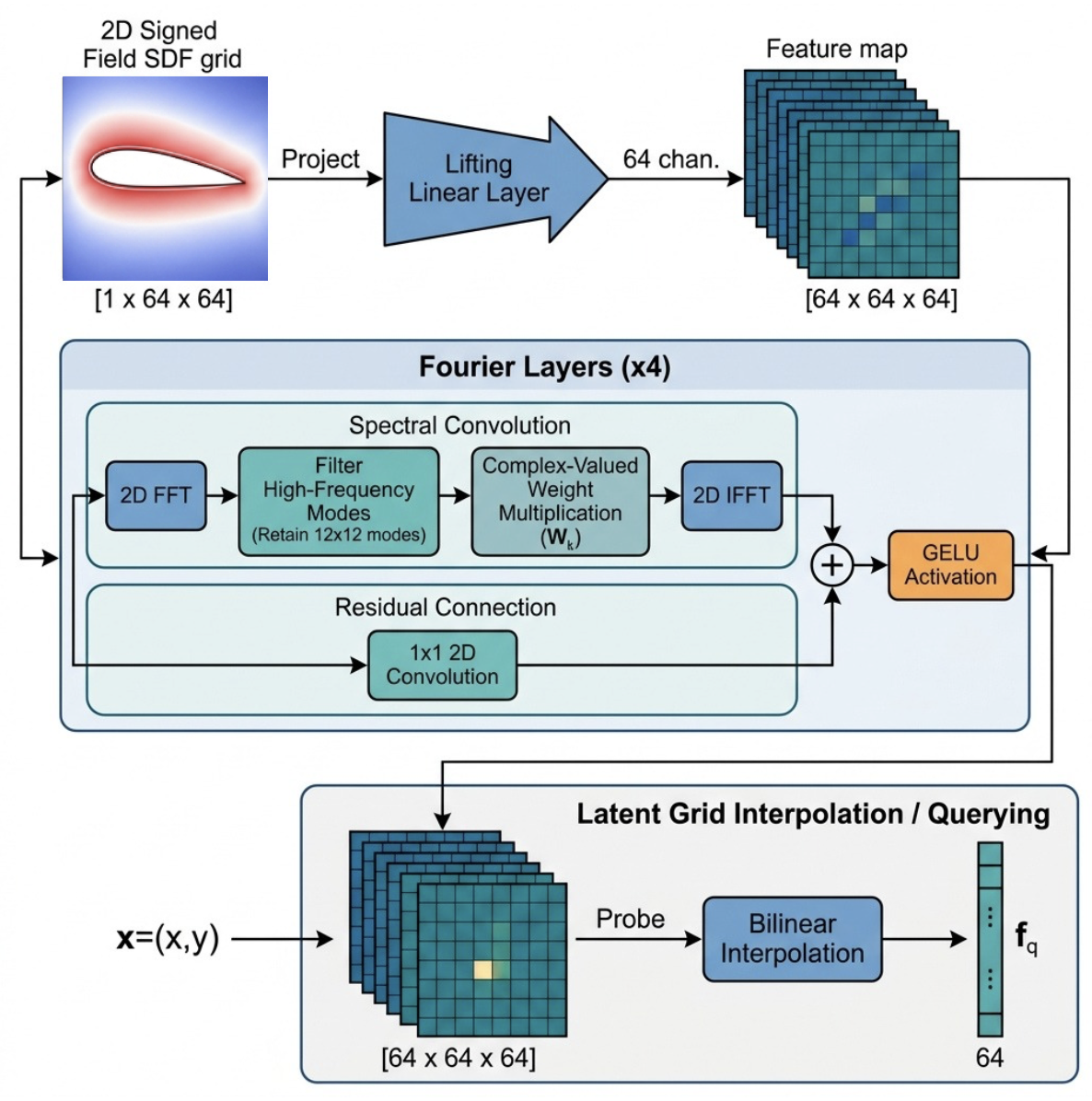}
    \caption{Architecture of the spectral geometry encoder.}
    \label{fig:rse}
\end{figure}

Instead of relying on a conventional CNN-based encoder that compresses the geometry into a single global latent vector, DD-RNO implements a 2D FNO~\cite{li2021fno} acting as a continuous Spectral Geometry Encoder. In traditional surrogate modeling, CNNs compress the input geometry into a single low-dimensional vector. While effective for simple classification, this compression discards critical local geometric details (such as the exact leading-edge nose radius or trailing-edge thickness) that dictate flow separation and lift/drag characteristics. By contrast, the FNO-based encoder preserves the spatial resolution of the geometry, producing a continuous, high-dimensional spatial feature map at the same $64\times64$ resolution. The detailed architecture of this encoder is illustrated in Figure~\ref{fig:rse}.

\subsubsection{Fourier Layers}

The FNO trunk consists of 4 spectral convolution layers. Each spectral layer performs a Fourier transform on the spatial feature map, filters out high-frequency modes (keeping the lowest $12\times12$ modes), applies a learnable complex-valued matrix multiplication to adjust these modes, and performs the inverse Fourier transform back to physical space. The layer uses Gaussian Error Linear Unit (GELU) activation functions:
\begin{equation}
\mathbf{h}^{(\ell+1)} = \text{GELU}(\mathcal{K}(\mathbf{h}^{(\ell)}) + \mathbf{W} \mathbf{h}^{(\ell)}),
\end{equation}
where $\mathcal{K}$ represents the spectral convolution, and $\mathbf{W}$ is a standard 2D convolution with a $1\times1$ kernel acting as a residual path. By operating in the frequency domain, the FNO layers efficiently capture global, low-frequency geometric patterns (such as overall airfoil thickness and camber distribution) and propagate this spatial context continuously across the entire computational domain. The width of the Fourier operator is set to 64 channels.

\subsubsection{Latent Grid Interpolation}

The output of the FNO trunk is a spatial feature map $\mathbf{Z} \in \mathbb{R}^{64 \times 64 \times 64}$. Because downstream decoders must predict flow variables at arbitrary continuous query coordinates $\mathbf{x} = [x, y]^\top$ (rather than just on a fixed grid), a method is required to extract the geometric features at these specific query points. This is achieved through Latent Grid Interpolation (LGI), which performs bilinear probing on the discrete feature map $\mathbf{Z}$:
\begin{equation}
\mathbf{f}_q = \text{BilinearProbe}(\mathbf{Z}, \mathbf{x}) \in \mathbb{R}^{64}.
\end{equation}
By mapping continuous coordinates directly onto the FNO feature grid, LGI serves as a local bridge that transforms grid-based global geometric features into a point-specific feature vector $\mathbf{f}_q$. This setup smoothly recovers localized sub-grid details without requiring complex, multi-scale grid pyramids.

\subsection{\textbf{Conditioning: Mapping Network and FiLM Modulation}}
\label{sec:conditioning}

\begin{figure}[!ht]
    \centering
    \includegraphics[width=\columnwidth]{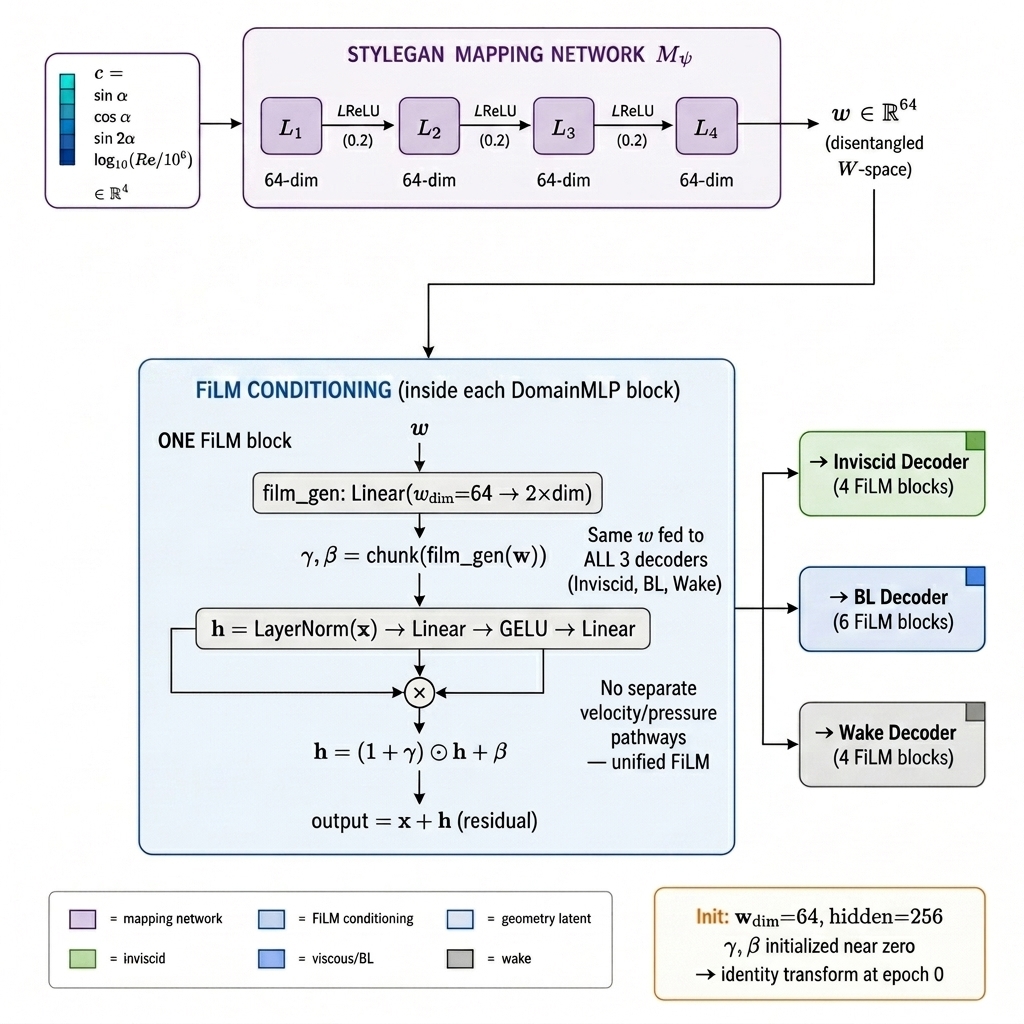}
    \caption{Mapping network and FiLM conditioning pathway.}
    \label{fig:conditioning}
\end{figure}

To make the network's behavior depend appropriately on flow conditions, a two-stage conditioning process is implemented. Figure~\ref{fig:conditioning} illustrates the complete conditioning pathway.

\subsubsection{Mapping Network}

Rather than feeding raw flow conditions (such as the angle of attack $\alpha$ or Reynolds number $\mathrm{Re}$) directly into the decoders alongside the spatial coordinates, a dedicated Mapping Network $M_\psi: \mathbb{R}^4 \to \mathbb{R}^{64}$ is employed. In standard networks, directly combining coordinate inputs with raw physical parameters forces the network to learn a highly non-linear, coupled space from scratch. The Mapping Network acts as a "disentangler" by processing the normalized flow conditions through four fully-connected layers with Leaky ReLU (LReLU) activations to produce a unified, high-dimensional $\mathbf{w}$-space latent vector:
\begin{equation}
\mathbf{w} = M_\psi(\mathbf{c}) = L_4(\text{LReLU}_{0.2}(L_3(\text{LReLU}_{0.2}(L_2(\text{LReLU}_{0.2}(L_1(\mathbf{c}))))))),
\end{equation}
where each $L_i$ denotes a fully-connected layer with $64$ hidden units. This latent representation $\mathbf{w}$ represents the "global flow state." Drawing inspiration from StyleGAN \cite{karras2019stylegan}, this formulation ensures that smooth interpolations within $\mathbf{w}$ map directly to continuous, physically realistic transitions in the predicted flow field (such as a smooth change in separation onset as angle of attack increases).

\subsubsection{FiLM-Conditioned Regional Field Decoders}

The global flow state latent $\mathbf{w}$ modulates the internal layers of each regional decoder using feature-wise linear modulation (FiLM) \cite{perez2018film}. Physically, instead of letting the decoders guess the flow conditions based on point coordinates, the latent state $\mathbf{w}$ acts as a global controller that scales (stretches) and shifts (offsets) the activation features at each network layer:
\begin{align}
(\mathbf{\gamma}^{(\ell)}, \mathbf{\beta}^{(\ell)}) &= \text{Linear}_{64 \to 2d}(\mathbf{w}), \\
\mathbf{h}^{(\ell)} &= \text{LayerNorm}(\mathbf{h}^{(\ell-1)}), \\
\mathbf{h}^{(\ell)} &= \text{Linear} \to \text{GELU} \to \text{Linear}(\mathbf{h}^{(\ell)}), \\
\mathbf{h}^{(\ell)} &= (\mathbf{1} + \mathbf{\gamma}^{(\ell)}) \odot \mathbf{h}^{(\ell)} + \mathbf{\beta}^{(\ell)}, \\
\mathbf{h}^{(\ell)} &= \mathbf{h}^{(\ell-1)} + \mathbf{h}^{(\ell)}.
\end{align}
Here, $\odot$ denotes the element-wise (Hadamard) product. The linear projection generates scale vectors $\mathbf{\gamma}^{(\ell)}$ and bias vectors $\mathbf{\beta}^{(\ell)}$ for each feature channel (where $\mathbf{1}$ represents the identity vector). The use of $(\mathbf{1} + \mathbf{\gamma}^{(\ell)})$ in the scale term ensures that the modulation begins as a near-identity perturbation during initialization, which stabilizes early training. This conditioning strategy allows the decoders to process spatial coordinate tokens while dynamically adjusting their mathematical functions based on the overall flow regime (e.g., automatically shifting boundary layer scaling when transitioning from low to high Reynolds numbers). A single, unified latent vector $\mathbf{w}$ is broadcast to all three decoders, eliminating the need for complex, decoupled parameter paths for different physical outputs.

\subsection{\textbf{Specialist Decoders and Physics-Guided Routing}}
\label{sec:routing}
The core structural innovation of DD-RNO is its deployment of three specialized regional decoders, each tailored to the distinct physical scales and gradient characteristics of a specific flow subregion. In aerodynamic applications, a single, uniform decoder typically struggles to represent the entire flow field because different zones exhibit radically different physics. For instance, the far-field inviscid flow varies slowly and smoothly, whereas the boundary layer next to the airfoil contains extremely sharp velocity gradients across a thin region. Driven by spectral bias, a single network will prioritize fitting the smooth far-field behavior and miss the critical near-wall dynamics entirely. To resolve this issue, the computational domain is partitioned into three distinct physical regions---Inviscid, Boundary Layer, and Wake---with a dedicated network assigned to each. When querying the flow field at any point, spatial coordinates are routed dynamically through these networks using differentiable, physics-guided gates.

\subsubsection{Decoder Token Representation}

Rather than passing raw spatial coordinates $\mathbf{x} = [x, y]^\top$ directly to the decoders, a continuous 101-dimensional input feature token $\mathbf{T}(\mathbf{x})$ is constructed at each query point:
\begin{equation}
\mathbf{T}(\mathbf{x}) = [\mathbf{\gamma}_{\text{MS}}(\mathbf{x}); \mathbf{f}_q; \Phi(\mathbf{x}); \hat{\mathbf{n}}(\mathbf{x})] \in \mathbb{R}^{101},
\end{equation}
where $\mathbf{\gamma}_{\text{MS}}(\mathbf{x}) \in \mathbb{R}^{34}$ is the multi-scale Fourier coordinate encoding (enabling the network to resolve sharp features), $\mathbf{f}_q \in \mathbb{R}^{64}$ represents the continuous global geometry features interpolated from the FNO trunk, $\Phi(\mathbf{x}) \in \mathbb{R}$ is the local SDF value (supplying the direct physical distance to the wall), and $\hat{\mathbf{n}}(\mathbf{x}) \in \mathbb{R}^2$ is the wall-normal unit vector at the closest surface point. This token maps a rich mixture of local coordinate positioning, global airfoil shape context, and wall proximity directly to the decoders, ensuring the networks have all the necessary geometric parameters to resolve the local flow.

\subsubsection{Decoder Specifications}

Each specialized decoder utilizes an MLP architecture embedded with the residual FiLM conditioning blocks described in Section~\ref{sec:conditioning}. Every decoder outputs a full 4-channel vector representing the velocity fields, pressure, and eddy viscosity: $\hat{\mathbf{q}} = [\hat{u}_x, \hat{u}_y, \hat{p}, \hat{\nu}_t]^\top$.
\begin{itemize}
    \item \textbf{Inviscid Decoder:} Designed to capture the smooth, low-frequency potential flow far from the airfoil. Because it doesn't need to resolve sharp near-wall gradients, it uses a lightweight architecture with 4 residual FiLM blocks (256 channels).
    \item \textbf{Boundary Layer Decoder:} Specialized exclusively in resolving the high-shear region next to the airfoil surface, where wall-normal gradients are steep and localized. To handle these high-frequency gradients, it uses a higher-capacity network with 6 residual FiLM blocks (256 channels).
    \item \textbf{Wake Decoder:} Focused on capturing downstream velocity deficits, pressure recovery, and recirculation zones behind the trailing edge. It consists of 4 residual FiLM blocks (256 channels).
\end{itemize}

\subsubsection{Physics-Adaptive Domain Routing}

Instead of using a complex, hard-to-interpret neural network to route query points, a simple physical rule based on turbulent boundary layer scaling is employed. The nominal boundary layer thickness $\delta_{\text{BL}}$ at a given Reynolds number is calculated dynamically using classical flat-plate turbulent scaling:
\begin{equation}
\delta_{\text{BL}}(\text{Re}) = c_{\text{BL}} \cdot \text{Re}^{-1/5},
\end{equation}
where $c_{\text{BL}} = 5.0$ is a scaling constant. Under classical flat-plate boundary layer theory \cite{schlichting2017boundary, white2006viscous}, the turbulent boundary layer thickness scales streamwise as $\delta(x) \approx 0.37 x \text{Re}_x^{-1/5}$; at the trailing edge ($x=c=1$), this yields a boundary layer thickness of $\delta(c) \approx 0.37 \text{Re}^{-0.2}$. We explicitly exclude the streamwise coordinate $x$ from the routing gate for two reasons. First, at the rounded airfoil leading edge ($x=0$), the flat-plate formula predicts a vanishing thickness ($\delta(0) = 0$). However, physical airfoils possess a leading-edge stagnation point and a severe suction peak with extremely high velocity and pressure gradients \cite{anderson2017fundamentals, abbott1959theory}. Shrinking the routing envelope to zero at the nose would route these critical near-wall gradients to the Inviscid decoder, causing immediate representation failure. Second, using a constant prefactor over the wall-normal distance $\Phi(\mathbf{x})$ forms a spatially uniform envelope that wraps around the entire airfoil contour. 

The prefactor $c_{\text{BL}} = 5.0$ is selected conservatively to account for high angles of attack, where strong adverse pressure gradients thicken the viscous shear layer well beyond flat-plate estimates. This ensures that the boundary layer decoder's spatial domain fully encloses all high-gradient near-wall structures, suction peaks, and incipient separation zones.

Unnormalized routing masks are then evaluated using smooth sigmoid transition functions:
\begin{align}
M_{\text{BL}}(\mathbf{x}) &= \sigma\left(\frac{\delta_{\text{BL}} - \Phi(\mathbf{x})}{0.1 \cdot \delta_{\text{BL}}}\right), \\
M_{\text{wake}}(\mathbf{x}) &= \sigma\left(\frac{x - x_{\text{wake}}}{0.05}\right),
\end{align}
where $\Phi(\mathbf{x})$ is the distance to the wall, and $x_{\text{wake}} = 1.05$ represents the wake gate starting location downstream of the trailing edge \cite{akolekar2019machine, pacciani2021assessment}. Physically, $M_{\text{BL}}(\mathbf{x})$ activates when a query point is close to the airfoil surface (within the boundary layer thickness $\delta_{\text{BL}}$), while $M_{\text{wake}}(\mathbf{x})$ activates when a query point is downstream of the airfoil. The asymmetry between the Reynolds-adaptive boundary layer gate and the geometrically fixed wake gate is physically motivated by the coordinate directions they govern. The boundary layer mask $M_{\text{BL}}$ controls the cross-stream transition (normal to the wall), where the physical scale of the viscous layer is extremely thin and highly sensitive to Reynolds number. In contrast, the wake mask $M_{\text{wake}}$ governs the stream-wise transition downstream of the trailing edge ($x_{\text{wake}} = 1.05$). The onset of the wake is geometrically fixed by the airfoil's physical trailing edge ($x = 1.0$), and the convective stream-wise transition length is primarily dictated by the chord scale rather than the Reynolds number. The fixed transition width of $0.05c$ ($5\%$ of the chord length) corresponds to the typical scale of trailing-edge pressure recovery and has been found to provide a sufficiently smooth transition across all test cases. The unnormalized inviscid mask is then defined as the remaining complement:
\begin{equation}
M_{\text{inv}}(\mathbf{x}) = (1.0 - M_{\text{BL}}(\mathbf{x})) \cdot (1.0 - M_{\text{wake}}(\mathbf{x})).
\end{equation}

To ensure smooth spatial transitions and prevent numerical artifacts at the boundaries between these regions, these masks are normalized to construct a partition of unity (so that the weights at any point sum to exactly 1.0):
\begin{equation}
G_k(\mathbf{x}) = \frac{M_k(\mathbf{x})}{\sum_i M_i(\mathbf{x})} \quad \text{for } k \in \{\text{inv}, \text{bl}, \text{wake}\}.
\end{equation}

The final predicted flow field is obtained by blending the outputs of the three specialized streams, weighted by their normalized routing gates:
\begin{equation}
\hat{\mathbf{q}}(\mathbf{x}) = G_{\text{inv}}(\mathbf{x}) \hat{\mathbf{q}}_{\text{inv}}(\mathbf{x}) + G_{\text{bl}}(\mathbf{x}) \hat{\mathbf{q}}_{\text{bl}}(\mathbf{x}) + G_{\text{wake}}(\mathbf{x}) \hat{\mathbf{q}}_{\text{wake}}(\mathbf{x}).
\end{equation}

This fully differentiable, physics-guided partition ensures smooth transitions across domain boundaries while allowing each sub-network to specialize in its respective flow regime.

\subsection{\textbf{Force Prediction via LCQ}}
\label{sec:lcq}

Instead of post-processing the volumetric field with an isolated pressure refinement head, the surface pressure distribution is extracted directly by evaluating the boundary layer decoder at the solid wall boundary. The boundary layer decoder is queried at a fixed set of $N_{\text{canon}} = 1024$ canonical surface coordinates $\mathbf{x}_{\text{surf}}$ where the signed distance field satisfies $\Phi(\mathbf{x}_{\text{surf}}) = 0$:
\begin{equation}
\mathbf{C}_p = \Pi_{p}\!\left(\text{BL\_Dec}\!\left([\mathbf{\gamma}_{\text{MS}}(\mathbf{x}_{\text{surf, norm}}); \mathbf{f}_s; 0; \hat{\mathbf{n}}(\mathbf{x}_{\text{surf}})], \mathbf{w}\right)\right),
\end{equation}
where $\Pi_{p}(\cdot)$ is a channel-selection operator that extracts the third (pressure) component from the decoder's 4-channel output vector $[u_x, u_y, p, \nu_t]^\top$. This predicted surface pressure coefficient vector, $\mathbf{C}_p \in \mathbb{R}^{N_{\text{canon}}}$, serves as the immediate input to the force prediction pipeline.

Predicting integrated aerodynamic forces (lift $C_L$ and drag $C_D$) from surrogate model outputs is traditionally executed via discrete geometric pressure integration over the airfoil surface:
\begin{equation}
C_L = \frac{1}{q_\infty c} \sum_j p_j (\mathbf{n}_j \cdot \hat{\mathbf{e}}_L) \ell_j, \quad C_D = \frac{1}{q_\infty c} \sum_j p_j (\mathbf{n}_j \cdot \hat{\mathbf{e}}_D) \ell_j,
\end{equation}
where $\hat{\mathbf{e}}_L = (-\sin\alpha, \cos\alpha)^\top$ and $\hat{\mathbf{e}}_D = (\cos\alpha, \sin\alpha)^\top$ define the lift and drag vector orientations relative to the chord line.
However, this classical discrete summation is highly sensitive to surface grid resolution and local interpolation errors. More importantly, standard pressure integration cannot capture skin friction (viscous) drag on its own, as that requires computing wall-normal velocity gradients ($\partial \mathbf{u}/\partial \mathbf{n} |_{\text{wall}}$)---an operation notorious for numerical instability when evaluated on neural network approximations of velocity fields.

\begin{figure*}[t]
    \centering
    \includegraphics[width=\textwidth]{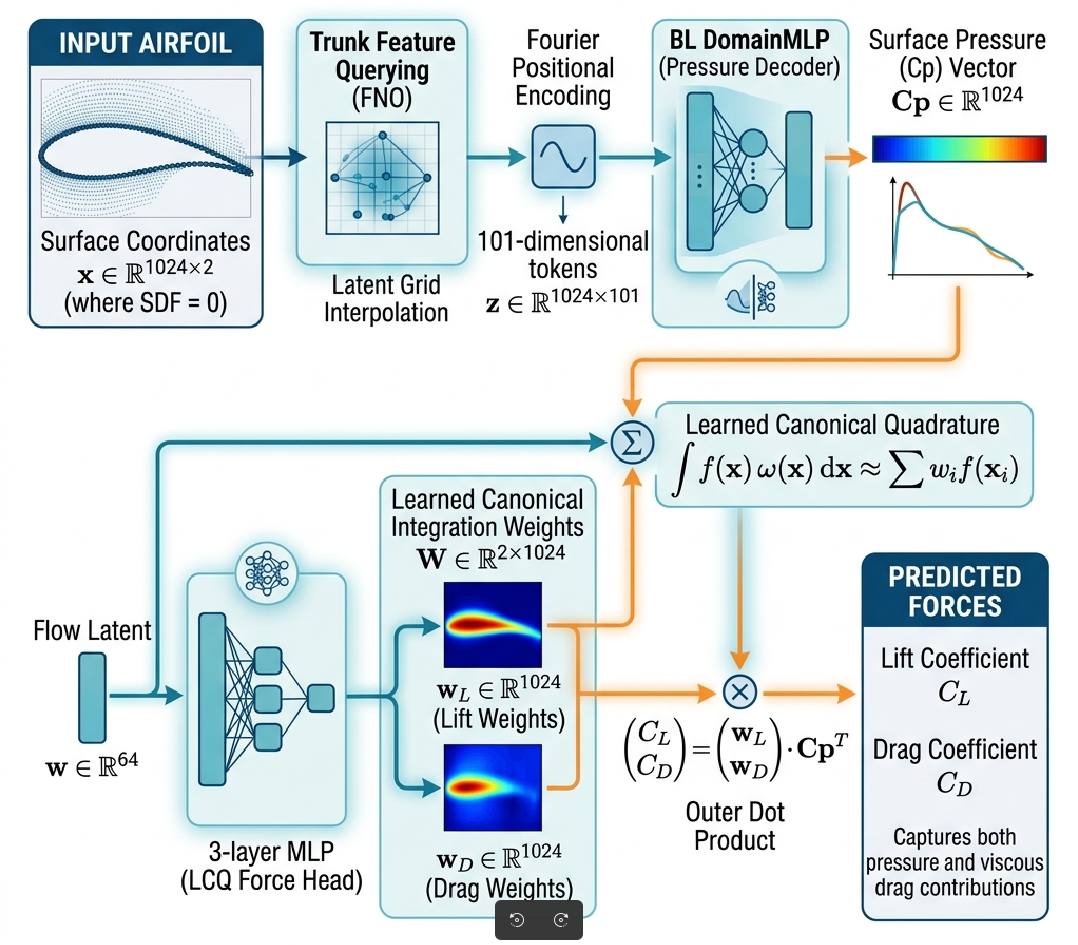}
    \caption{LCQ surface pressure and force prediction pipeline. }
    \label{fig:lcq_pipeline}
\end{figure*}

The proposed LCQ framework bypasses these geometric and numerical limitations by reformulating force integration as a learned, flow-conditioned inner product. A 3-layer MLP maps the global flow condition latent $\mathbf{w}$ to a set of dynamic canonical quadrature weights:
\begin{equation}
\mathbf{W}_{\text{canon}} = f_{\text{LCQ}}(\mathbf{w}) \in \mathbb{R}^{2 \times N_{\text{canon}}}.
\end{equation}
The total integrated aerodynamic forces are then computed via the direct dot product:
\begin{equation}
\begin{bmatrix} \hat{C}_L \\ \hat{C}_D \end{bmatrix} = \sum_{j=1}^{N_{\text{canon}}} \mathbf{W}_{\text{canon}, j}(\mathbf{w}) \cdot C_p^{(j)}.
\end{equation}
This formulation delivers three key advantages: (1) it lets the network learn optimal integration weights that adapt dynamically to changing flow regimes, (2) it naturally compensates for spatial discretization errors across the canonical boundary representation, and (3) it statistically reconstructs total aerodynamic drag directly from the surface pressure distribution and global latent state. By capturing both pressure and viscous components simultaneously, it isolates a subtle yet statistically significant geometry-dependent viscous drag signal beyond flow-condition confounds. Consequently, this design eliminates the need for standalone viscous drag correction factors or unstable wall-normal gradient calculations. The complete end-to-end pipeline is illustrated in Figure~\ref{fig:lcq_pipeline}.

While data-driven integration and learned quadrature methods (such as the Empirical Cubature Method \cite{hernandez2017empirical} or Energy-Conserving Sampling and Weighting\cite{farhat2014energy}) are established in computational physics for hyper-reduction of reduced-order models, LCQ differs from these classical formulations in three fundamental aspects. First, traditional hyper-reduction quadrature constructs a static set of weights and nodes offline to minimize PDE residuals. In contrast, LCQ dynamically predicts the integration weights $\mathbf{W}_{\text{canon}}(\mathbf{w})$ at inference time as a function of the global flow latent state, enabling the integration to adapt to shifting flow regimes (such as transitioning or separated flow states). Second, rather than acting purely as a numerical approximator of a known integral, LCQ is trained end-to-end to implicitly reconstruct the total force—partially capturing the unmodeled skin friction (viscous) drag directly from pressure distributions without querying velocity fields or computing unstable wall-normal gradients. Finally, it integrates continuous fields queried over a standardized canonical coordinate system, ensuring geometric flexibility across arbitrary airfoil profiles.

To verify that the LCQ module partially recovers true physical viscous drag rather than simply acting as an arbitrary fix for local pressure errors, a post-hoc decomposition analysis is conducted. In aerodynamics, total drag is the sum of pressure drag and viscous drag ($C_D = C_{Dp} + C_{Df}$). We define the model's implicit drag correction as $\Delta \hat{C}_D = \hat{C}_D - C_{Dp,\text{geom}}(\hat{\mathbf{C}}_p)$, where $\hat{C}_D$ is the total drag predicted by LCQ, and $C_{Dp,\text{geom}}(\hat{\mathbf{C}}_p)$ is the pressure drag obtained via standard geometric integration over the predicted surface pressure distribution $\hat{\mathbf{C}}_p$.  The correlation between this implicit correction $\Delta \hat{C}_D$ and the ground-truth viscous drag $C_{Df}$ is evaluated from the RANS simulations across the test split ($N = 200$). Because both terms depend heavily on operating conditions, a partial correlation analysis is run to control for the confounding effects of angle of attack ($\alpha$) and Reynolds number ($\log_{10}(\text{Re})$). This revealed a statistically significant positive relationship, yielding a partial Spearman correlation of $\rho = 0.3365$ ($p_{\text{val}} < 0.001$, partial $R^2 \approx 0.081$).  While highly significant, the modest $R^2$ value indicates that geometry-specific features explain only a small fraction ($\approx 8.1\%$) of the residual viscous drag variance once flow conditions are controlled for. This implies that while LCQ does capture some geometry-dependent viscous signals (such as thickness-induced skin friction variations), a lot of the uncorrected correlation is driven by flow-condition variables ($\alpha$ and $\text{Re}$) rather than a complete, standalone recovery of viscous physics.  Additionally, the correlation between the learned LCQ weights and traditional analytical integration weights is virtually zero ($\rho \approx -0.02$, with a $95\%$ confidence interval of $[-0.16, 0.12]$). This near-zero correlation shows that the network is not merely mimicking standard geometric pressure integration (which would only yield the pressure drag $C_{Dp}$). Instead, it learns an alternative mapping that combines surface pressure signatures with the global flow latent state $\mathbf{w}$ to achieve statistical error cancellation while implicitly capturing viscous drag.

\subsection{\textbf{Training and Losses}}
\label{sec:training}

\subsubsection{Loss Functions}

The total loss objective is optimized to balance field reconstruction with integrated force predictions:
\begin{equation}
\mathcal{L}_{\text{total}} = \lambda_{\text{field}} \mathcal{L}_{\text{field}} + \lambda_{\text{force}} \mathcal{L}_{\text{force}}.
\end{equation}

The field loss evaluates a weighted MSE across the full suite of flow variables, defined as $\mathbf{q} = [u_x, u_y, p, \nu_t]^\top$. To prioritize training resolution inside the boundary layer where downstream force computations exhibit extreme sensitivity, an exponential weighting function based on the SDF is applied:
\begin{equation}
W(\mathbf{x}) = \exp(-15.0 \cdot |\Phi(\mathbf{x})|) + 0.1,
\end{equation}
\begin{align}
\mathcal{L}_{\text{field}} = \frac{1}{BQ} \sum_{b,q} W(\mathbf{x}_{b,q}) \Big[ &(\hat{u}_x - u_x)^2 + (\hat{u}_y - u_y)^2 \nonumber \\
&+ (\hat{p} - p)^2 + (\hat{\nu}_t - \nu_t)^2 \Big],
\end{align}
where $B$ represents the batch size and $Q$ denotes the number of sampled query coordinates.

The corresponding force loss calculates the MSE over the Z-score standardized lift and drag coefficients:
\begin{equation}
\mathcal{L}_{\text{force}} = \frac{1}{B} \sum_b \left[ (\hat{C}_{L,b} - C_{L,b}^{\text{CFD}})^2 + (\hat{C}_{D,b} - C_{D,b}^{\text{CFD}})^2 \right].
\end{equation}

\subsubsection{Loss Balancing}

\begin{figure}[t]
    \centering
    \includegraphics[width=\columnwidth]{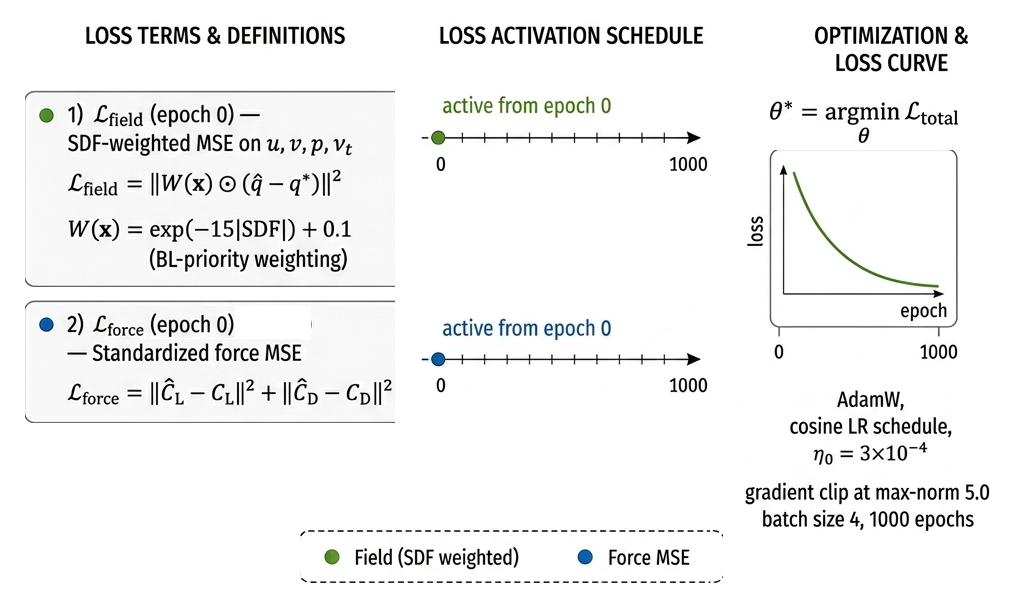}
    \caption{Training loss structure and optimization schedule.}
    \label{fig:loss}
\end{figure}

While the codebase supports adaptive multi-task uncertainty weighting \cite{kendall2018multi} as an alternate runtime configuration, empirical testing demonstrated that fixed, static loss weights provide robust and highly reproducible convergence. For all final publication benchmarks, static weights are set to $\lambda_{\text{field}} = 100.0$ and $\lambda_{\text{force}} = 10.0$. Figure~\ref{fig:loss} outlines the structural breakdown of the training losses alongside the optimization schedule.

\subsubsection{Optimization}

Model optimization is conducted using the AdamW optimizer with hyperparameters set to $\beta_1 = 0.9$, $\beta_2 = 0.999$, $\epsilon = 10^{-8}$, an initial learning rate of $\eta_0 = 3\times 10^{-4}$, and a decoupled weight decay coefficient of $\lambda_{\text{wd}} = 10^{-4}$. The learning rate is adjusted via a cosine annealing schedule \cite{loshchilov2016sgdr}:
\begin{equation}
\eta(e) = \eta_{\min} + \frac{1}{2}(\eta_0 - \eta_{\min})\left(1 + \cos\left(\frac{\pi e}{E_{\max}}\right)\right),
\end{equation}
where $E_{\text{max}} = 1000$ and $\eta_{\min} \approx 0$. Gradients are scaled using a maximum norm threshold of 5.0. Training is executed with a batch size of 4, yielding approximately 7.5M total trainable parameters.

\subsubsection{Reproducibility}

All numerical experiments were conducted using PyTorch 1.13 and CUDA 11.7. The AirfRANS dataset (v1.0)~\cite{bonnet2023airfrans} was evaluated strictly according to its standardized testing protocol. Model training was performed on an NVIDIA RTX A6000 GPU with a batch size of 4, requiring approximately $4.5$ hours of training time to optimize the 7.5M trainable parameters. Inference benchmarks were conducted on an NVIDIA RTX 2050 GPU, yielding an average execution time of $144$ ms per sample.

\section{Results \& Discussions}
\label{sec:results}

In this section, all numerical evaluations have been conducted on the standardized AirfRANS benchmark dataset \cite{bonnet2023airfrans}. Model predictions have been compared directly against established baseline architectures evaluated on identical data splits on the AirfRANS dataset: MLPs, GraphSAGE \cite{hamilton2017graphsage}, and Graph U-Nets \cite{gao2019graphunet}.
Evaluation is performed across three standardized benchmark tasks from AirfRANS \cite{bonnet2023airfrans}:
\begin{itemize}
    \item \textbf{Full} -- The standard split comprising $N_{\text{train}} = 800$ training samples and $N_{\text{test}} = 200$ test samples (1,000 total RANS simulations), spanning angles of attack $\alpha \in [-5^\circ, 25^\circ]$ and Reynolds numbers $\text{Re} \in [2\times10^6, 6\times10^6]$.
    \item \textbf{Reynolds out-of-distribution (OOD)} -- OOD  extrapolation comprising $N_{\text{train}} = 800$ training samples restricted to the intermediate band $\text{Re} \in [3.02\times10^6, 5.03\times10^6]$, and $N_{\text{test}} = 200$ test samples evaluating performance on unseen lower ($\text{Re} \in [2.02\times10^6, 3.02\times10^6]$) and upper ($\text{Re} \in [5.03\times10^6, 6.04\times10^6]$) Reynolds bands.
    \item \textbf{Angle of Attack (AoA) OOD} -- OOD extrapolation comprising $N_{\text{train}} = 800$ training samples restricted to moderate angles $\alpha \in [-2.48^\circ, 12.45^\circ]$, and $N_{\text{test}} = 200$ test samples evaluating extreme negative ($\alpha < -2.48^\circ$) and high positive ($\alpha > 12.45^\circ$) angles of attack.
\end{itemize}
All field error metrics are evaluated directly at native CFD mesh nodes without spatial resampling.

\begin{figure*}[t]
    \centering
    \begin{subfigure}{\textwidth}
        \centering
        \includegraphics[width=\textwidth]{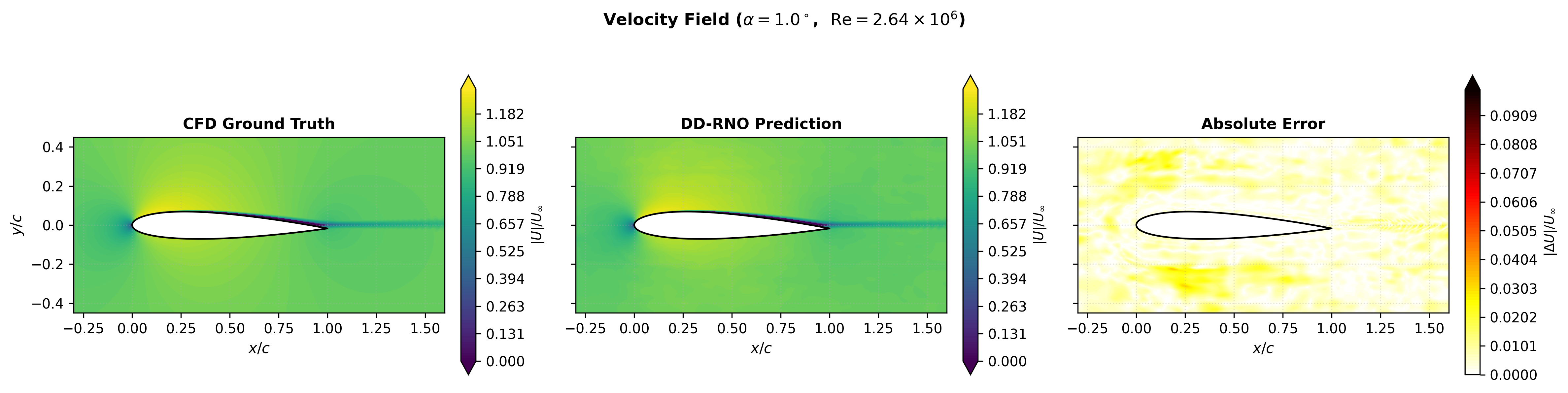}
        \caption{Velocity magnitude at $\alpha=1.1^\circ$ ($\text{Re} \approx 2 \times 10^6$): CFD ground truth (left), DD-RNO prediction (center), absolute error field (right).}
        \label{fig:vis_aoa1}
    \end{subfigure}
    \vskip 1em
    \begin{subfigure}{\textwidth}
        \centering
        \includegraphics[width=\textwidth]{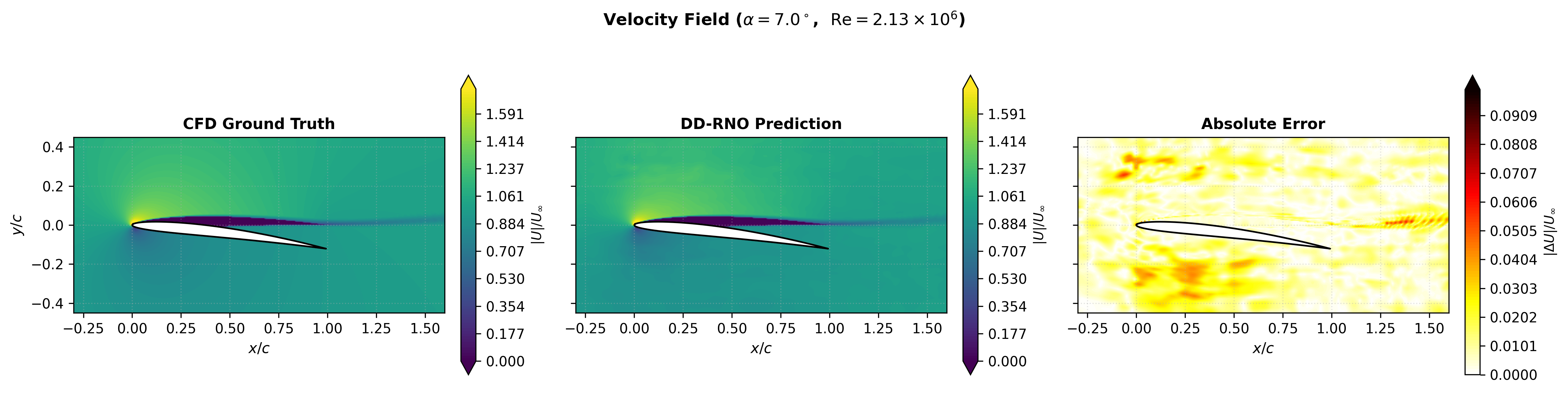}
        \caption{Velocity magnitude at $\alpha=7.4^\circ$ ($\text{Re} \approx 2 \times 10^6$): CFD ground truth (left), DD-RNO prediction (center), absolute error field (right).}
        \label{fig:vis_aoa7}
    \end{subfigure}
    \vskip 1em
    \begin{subfigure}{\textwidth}
        \centering
        \includegraphics[width=\textwidth]{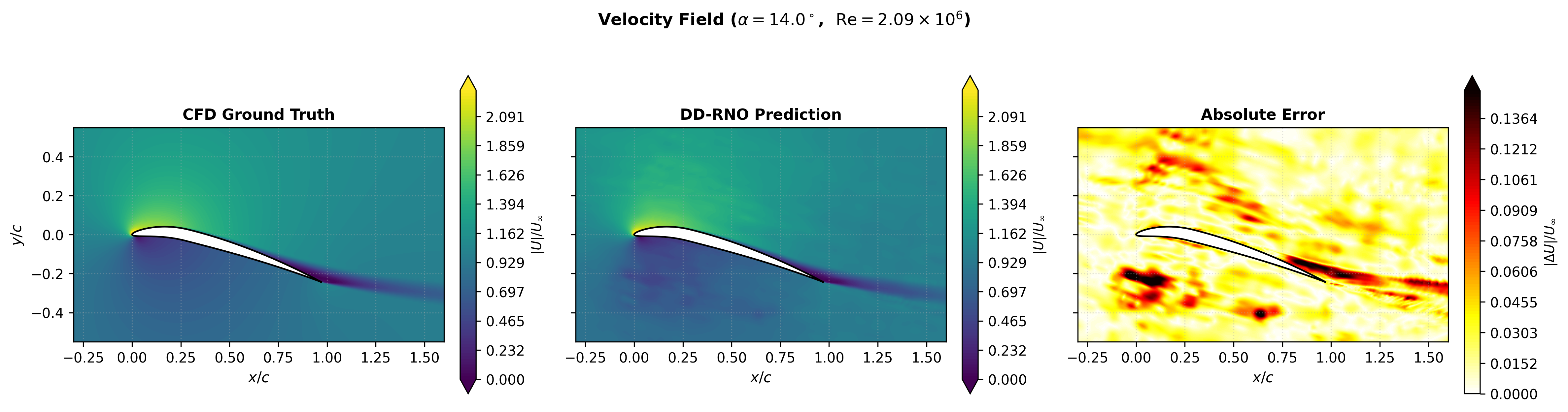}
        \caption{Velocity magnitude at $\alpha=14.1^\circ$ ($\text{Re} \approx 2 \times 10^6$): CFD ground truth (left), DD-RNO prediction (center), absolute error field (right).}
        \label{fig:vis_aoa14}
    \end{subfigure}
    \caption{Qualitative comparison of DD-RNO flow field predictions against OpenFOAM CFD ground truth for representative angles of attack from the Full test set at $\text{Re} \approx 2 \times 10^6$}
    \label{fig:fields}
\end{figure*}

\subsection{\textbf{Field Reconstruction Accuracy}}
\label{sec:results_field}

\subsubsection{Quantitative Benchmark Metrics}

Table~\ref{tab:field_mse} compiles field reconstruction MSE across all benchmarked methods. On the standard Full task, DD-RNO achieves substantial accuracy gains over all established baselines. Compared to the best-performing baseline model -- MLP, DD-RNO reduces velocity MSE by $17\times$ in $u_x$ and $12\times$ in $u_y$. Volumetric pressure MSE is reduced by $3.4\times$, while surface pressure error is reduced by $1.7\times$.
On the Reynolds OOD task, performance improvements are even larger: DD-RNO achieves a $23\times$ reduction in $u_x$ MSE compared to the best baseline MLP, an $11\times$ reduction in $u_y$ MSE, and a $14\times$ reduction in volumetric pressure error. This strong performance highlights the physical generalization of the Reynolds-adaptive boundary layer routing mechanism ($\delta_{\text{BL}} \propto \text{Re}^{-1/5}$).
For the AoA OOD extrapolation task, DD-RNO maintains superior velocity field accuracy, reducing $u_x$ MSE by $3.1\times$ and $u_y$ MSE by $4.3\times$ compared to the best baseline models.

\subsubsection{Flow Field Visualizations and Spatial Error Interpretation} Figure~\ref{fig:fields} provides a qualitative visual comparison of DD-RNO velocity predictions against OpenFOAM CFD ground truth across low ($\alpha=1.1^\circ$), moderate ($\alpha=7.4^\circ$), and high ($\alpha=14.1^\circ$) angles of attack at $\text{Re} \approx 2 \times 10^6$. At low and moderate angles ($\alpha = 1.1^\circ$ and $7.4^\circ$), the predicted velocity magnitude fields exhibit near-perfect visual and quantitative alignment with CFD ground truth, with absolute local velocity errors remaining below $0.02 \, U_\infty$ across both the attached boundary layer and the trailing wake. At high angle of attack ($\alpha = 14.1^\circ$), absolute error increases locally within the downstream wake region ($x/c > 1.2$). This spatial error localization occurs because steep velocity deficits and expanding trailing-edge shear layers amplify small spatial phase shifts in the coordinate grid, whereas the near-wall boundary layer, leading-edge stagnation point, and suction peak over the upper airfoil surface remain accurately resolved with minimal error.

\begin{figure}[t]
    \centering
    \includegraphics[width=0.75\linewidth]{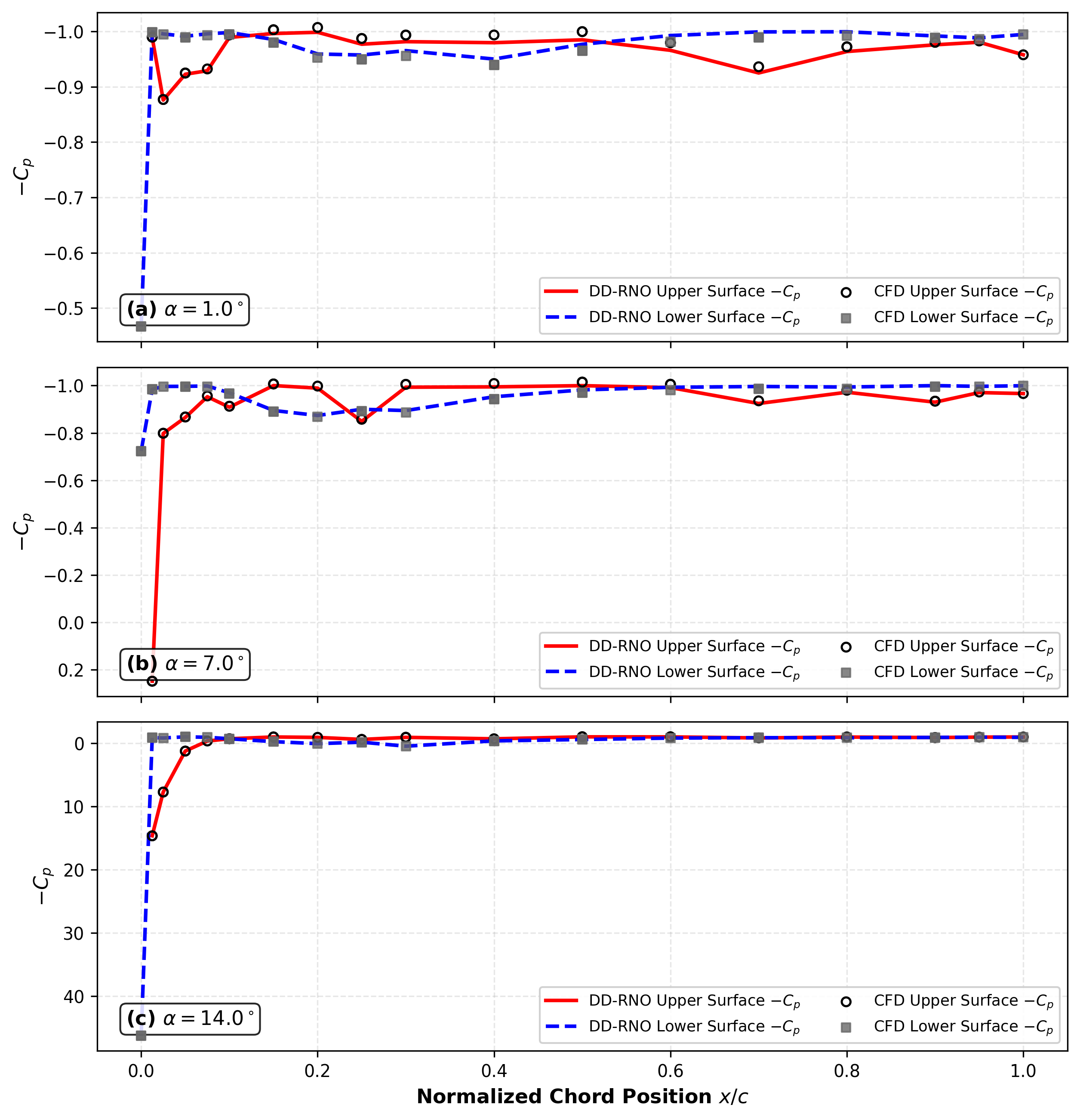}
    \caption{Surface pressure distribution $-C_p(x/c)$ along the chord of a NACA 0012 airfoil at $\text{Re} = 3.0\times 10^6$ for (a) low ($\alpha = 1.0^\circ$), (b) moderate ($\alpha = 7.0^\circ$), and (c) high ($\alpha = 14.0^\circ$) angles of attack}
    \label{fig:pressure_profiles}
\end{figure}

\subsubsection{Surface Pressure Distribution Overlay} As an intermediate operational step prior to force quadrature, the boundary layer decoder directly outputs surface pressure distributions $C_p(x/c)$. Figure~\ref{fig:pressure_profiles} overlays predicted surface pressure profiles against CFD ground-truth distributions for NACA 0012 across low ($\alpha=1.0^\circ$), moderate ($\alpha=7.0^\circ$), and high ($\alpha=14.0^\circ$) angles of attack. DD-RNO captures the sharp suction peak at the leading edge ($x/c < 0.05$) and the adverse pressure gradient recovery along the upper surface with near-zero discrepancy, confirming that the boundary layer decoder captures physical pressure distributions prior to integration.

\begin{table*}[t]
\centering
\caption{Field Reconstruction MSE ($\times 10^{-2}$). Lower is better. Baseline values from Bonnet et al. \cite{bonnet2023airfrans}.}
\label{tab:field_mse}
\begin{tabular}{@{}lcccc@{}}
\toprule
\textbf{Field / Metric} & \textbf{MLP} & \textbf{GraphSAGE} & \textbf{Graph U-Net} & \textbf{DD-RNO} \\
\midrule
\multicolumn{5}{c}{\textbf{Full Task} ($N_{\text{train}}=800, N_{\text{test}}=200$)} \\
\midrule
$u_x$ & $1.58$ & $1.72$ & $1.68$ & $\mathbf{0.091}$ \\
$u_y$ & $1.41$ & $1.53$ & $1.45$ & $\mathbf{0.114}$ \\
$p$   & $0.76$ & $0.81$ & $0.79$ & $\mathbf{0.225}$ \\
$p_s$ & $1.44$ & $1.52$ & $1.48$ & $\mathbf{0.854}$ \\
\midrule
\multicolumn{5}{c}{\textbf{Reynolds OOD Task} ($N_{\text{train}}=800, N_{\text{test}}=200$)} \\
\midrule
$u_x$ & $10.2$ & $10.8$ & $10.5$ & $\mathbf{0.448}$ \\
$u_y$ & $5.72$ & $6.01$ & $5.89$ & $\mathbf{0.541}$ \\
$p$   & $5.38$ & $5.72$ & $5.55$ & $\mathbf{0.388}$ \\
$p_s$ & $11.5$ & $12.1$ & $11.8$ & $\mathbf{1.266}$ \\
\midrule
\multicolumn{5}{c}{\textbf{AoA OOD Task} ($N_{\text{train}}=800, N_{\text{test}}=200$)} \\
\midrule
$u_x$ & $5.64$ & $6.12$ & $5.91$ & $\mathbf{1.817}$ \\
$u_y$ & $8.88$ & $9.24$ & $9.01$ & $\mathbf{2.045}$ \\
$p$   & $9.79$ & $10.12$ & $9.95$ & $11.337$ \\
$p_s$ & $23.6$ & $24.1$ & $23.8$ & $36.766$ \\
\bottomrule
\end{tabular}
\end{table*}

\subsubsection{Out-of-Distribution Validation at Low Reynolds Number ($\text{Re}=1.0\times 10^6$)}
\label{sec:results_1m}

To evaluate surrogate stability under physical extrapolation below the training domain threshold of $\text{Re}_{\min} = 2\times 10^6$, DD-RNO was benchmarked directly against a  OpenFOAM RANS CFD simulation of a NACA 0012 airfoil at $\text{Re} = 1.0\times 10^6$ and $\alpha = 0.0^\circ$.
 DD-RNO demonstrates strong near-field and far-field generalization under OOD Reynolds-number extrapolation. The surrogate accurately reproduces key aerodynamic flow features, including the leading-edge stagnation region ($|U|/U_\infty \approx 0$), the symmetric surface acceleration and deceleration patterns, and the downstream wake structure observed in the CFD solution.

\begin{figure*}[t]
    \centering
    \includegraphics[width=0.92\textwidth]{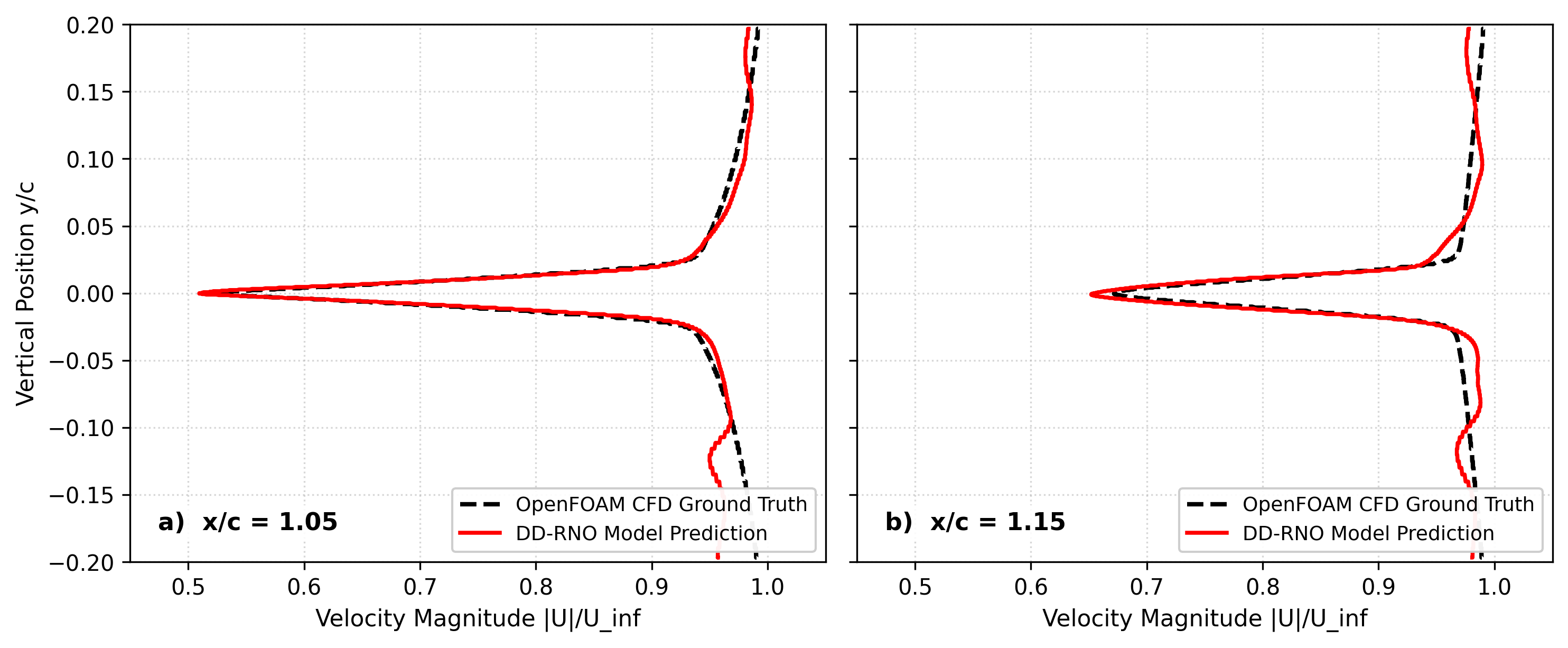}
    \caption{Out-of-distribution trailing-edge wake velocity profile comparison ($|U|/U_\infty$) for the NACA 0012 airfoil at $\text{Re} = 1.0\times 10^6$ and $\alpha = 0.0^\circ$ at (a) $x/c = 1.05$ and (b) $x/c = 1.15$. 
    }
    \label{fig:wake_profiles}
\end{figure*}

 Figure~\ref{fig:wake_profiles} presents wake velocity profiles extracted at $x/c = 1.05$ and $x/c = 1.15$. At both locations, DD-RNO accurately captures the wake velocity deficit, including the minimum velocity at the wake centerline ($y/c = 0$) and the gradual recovery toward freestream conditions. The close agreement with the OpenFOAM reference solution is particularly noteworthy, as wake regions contain sharp velocity gradients and thin shear layers that are traditionally challenging for surrogate models to resolve. Accurate prediction of the wake deficit is critical because it directly influences aerodynamic force estimation, downstream flow interactions, and loss calculations. Minor discrepancies in the freestream region are attributed to the adaptive decoder allocation strategy, which concentrates representational capacity in the wake and boundary-layer regions where flow complexity is greatest, while employing fewer decoders in relatively uniform freestream zones.

Across the entire physical flow domain, DD-RNO achieves a velocity MAE of $0.0382$ ($3.82\%$ relative error), demonstrating robust predictive accuracy under out-of-distribution Reynolds-number extrapolation. These results indicate that the surrogate successfully preserves the underlying flow physics while maintaining stable, non-divergent predictions beyond the Reynolds-number range encountered during training. 


\subsection{\textbf{Force Coefficient Accuracy}}
\label{sec:results_forces}

\subsubsection{Integrated Force Benchmark Evaluation}

Table~\ref{tab:forces} details integrated force coefficient errors and Spearman rank correlation coefficients ($\rho$). For the standard Full task, DD-RNO reduces relative drag error from $6.178\%$ (best baseline MLP) down to $1.091\%$ (a $5.6\times$ error reduction), and relative lift error from $14.8\%$ (best baseline GraphSAGE) down to $3.276\%$ (a $4.5\times$ error reduction). Crucially, the Spearman drag rank correlation increases from $\rho = 0.250$ for the best baseline model up to $\rho = 0.997$ for DD-RNO. This high rank correlation demonstrates that DD-RNO preserves the true aerodynamic performance ranking across arbitrary airfoil geometries.
\begin{table*}[t]
\centering
\caption{Force Coefficient Errors and Spearman Rank Correlations ($\rho$). Lower relative error is better; higher $\rho$ is better. Baseline values from Bonnet et al. \cite{bonnet2023airfrans}.}
\label{tab:forces}
\begin{tabular}{@{}lcccc@{}}
\toprule
\textbf{Metric} & \textbf{MLP} & \textbf{GraphSAGE} & \textbf{Graph U-Net} & \textbf{DD-RNO} \\
\midrule
\multicolumn{5}{c}{\textbf{Full Task} ($N_{\text{train}}=800, N_{\text{test}}=200$)} \\
\midrule
$C_D$ Rel. Error (\%) & $6.178$ & $7.366$ & $13.320$ & $\mathbf{1.091}$ \\
$C_L$ Rel. Error (\%) & $21.1$ & $14.8$ & $16.8$ & $\mathbf{3.276}$ \\
$\rho_D$ (Spearman) & $0.250$ & $0.194$ & $0.092$ & $\mathbf{0.9973}$ \\
$\rho_L$ (Spearman) & $0.993$ & $0.996$ & $0.995$ & $\mathbf{0.9992}$ \\
\midrule
\multicolumn{5}{c}{\textbf{Reynolds OOD Task} ($N_{\text{train}}=800, N_{\text{test}}=200$)} \\
\midrule
$C_D$ Rel. Error (\%) & $8.293$ & $12.794$ & $18.103$ & $\mathbf{2.728}$ \\
$C_L$ Rel. Error (\%) & $62.1$ & $43.3$ & $46.6$ & $\mathbf{8.585}$ \\
$\rho_D$ (Spearman) & $0.157$ & $0.039$ & $0.192$ & $\mathbf{0.9906}$ \\
$\rho_L$ (Spearman) & $0.958$ & $0.971$ & $0.964$ & $\mathbf{0.9974}$ \\
\midrule
\multicolumn{5}{c}{\textbf{AoA OOD Task} ($N_{\text{train}}=800, N_{\text{test}}=200$)} \\
\midrule
$C_D$ Rel. Error (\%) & $\mathbf{4.355}$ & $6.047$ & $9.814$ & $7.119$ \\
$C_L$ Rel. Error (\%) & $41.3$ & $\mathbf{25.4}$ & $37.6$ & $26.352$ \\
$\rho_D$ (Spearman) & $0.347$ & $0.525$ & $0.552$ & $\mathbf{0.9392}$ \\
$\rho_L$ (Spearman) & $0.957$ & $0.989$ & $0.982$ & $\mathbf{0.9903}$ \\
\bottomrule
\end{tabular}
\end{table*}
Under the Reynolds OOD configuration, relative drag error is reduced from $8.293\%$ (MLP) down to $2.728\%$, while relative lift error drops from $43.3\%$ (GraphSAGE) to $8.585\%$. Spearman drag correlation remains near-unity ($\rho = 0.991$ versus baseline $\rho = 0.157$).
On the AoA OOD task, DD-RNO achieves a Spearman drag correlation of $\rho = 0.939$ (compared to $\rho = 0.552$ for the best baseline Graph U-Net), confirming robust aerodynamic trend tracking even under out-of-distribution physical extrapolation.

\begin{figure*}[t]
    \centering
    \includegraphics[width=0.95\textwidth]{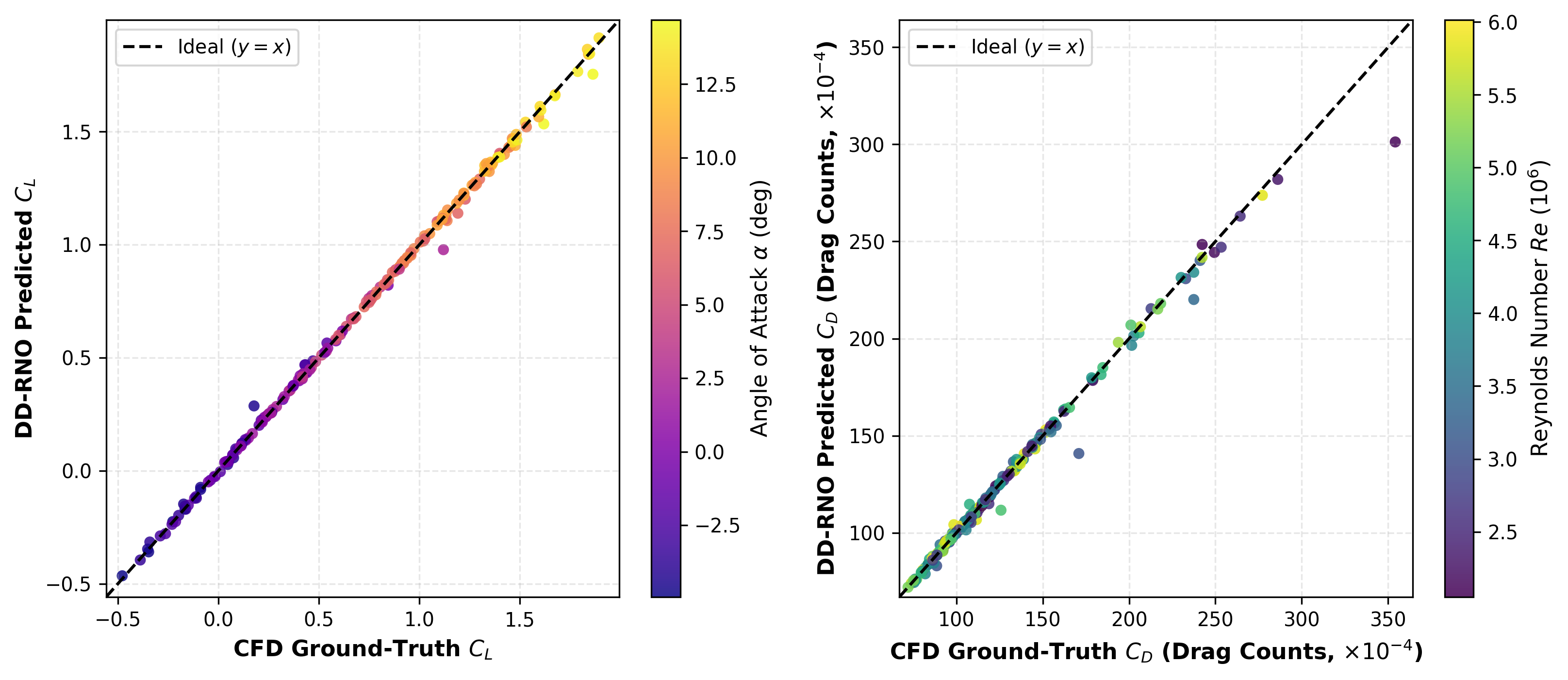}
    \caption{Parity and scatter comparison of predicted aerodynamic force coefficients against CFD ground truth across the 200 test samples from the Full task}
    \label{fig:parity_plots}
\end{figure*}

\subsubsection{Force Prediction Parity Analysis} Figure~\ref{fig:parity_plots} illustrates the parity scatter plots of predicted versus CFD ground-truth force coefficients across all 200 test samples from the Full task. The lift coefficient $C_L$ (left panel) exhibits exceptional agreement ($R^2 = 0.9994$, relative error $= 3.72\%$), with samples closely adhering to the $y=x$ ideal line across the entire range of angles of attack ($\alpha \in [-5^\circ, 25^\circ]$). Similarly, drag coefficient $C_D$ (right panel) demonstrates near-perfect alignment in drag counts ($R^2 = 0.9975$, relative error $= 1.091\%$, Spearman $\rho = 0.9975$), validating that LCQ maintains high accuracy without systematic drag offsets.

\subsubsection{Parametric Aerodynamic Trend Validation} 
\begin{figure*}[t]
    \centering
    \includegraphics[width=0.92\textwidth]{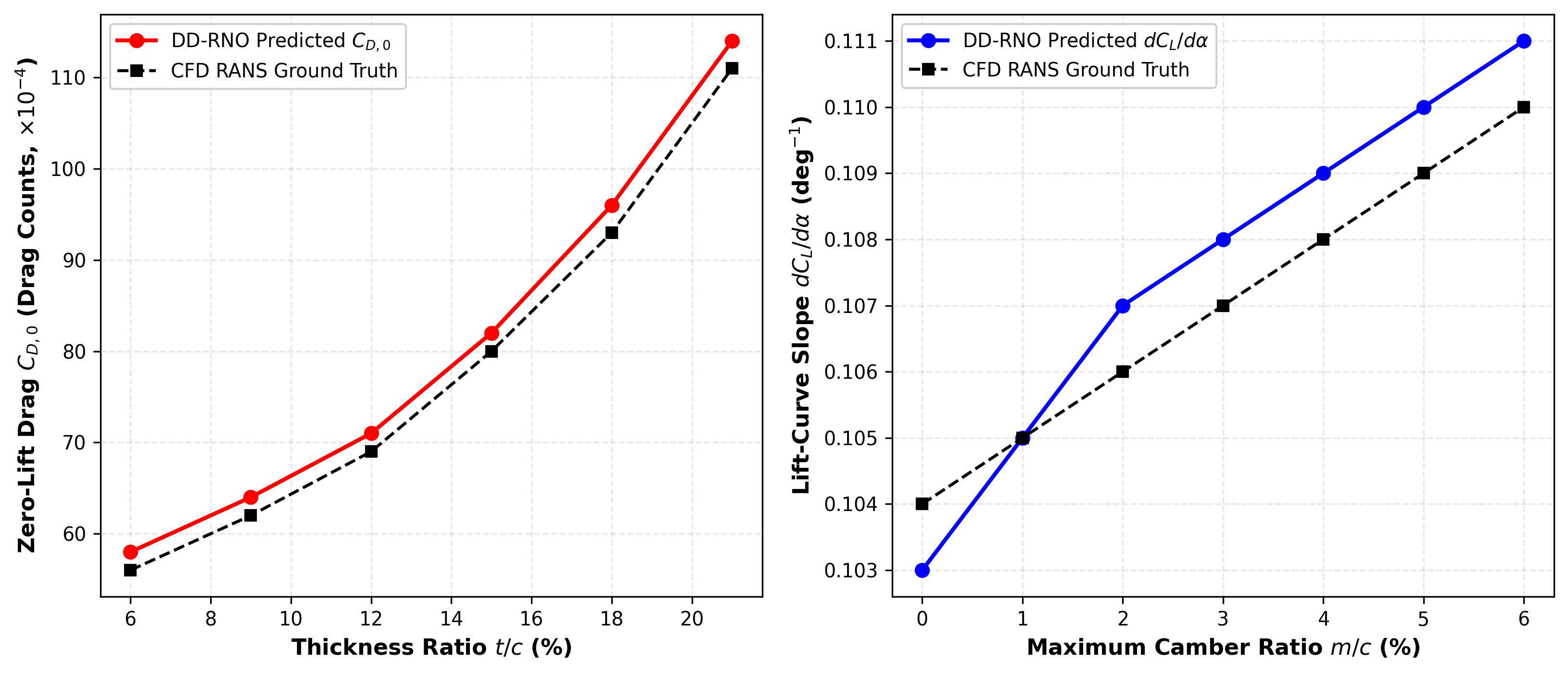}
    \caption{Aerodynamic trend validation across NACA-family airfoils}
    \label{fig:naca_trends}
\end{figure*}

To confirm that DD-RNO captures physical aerodynamic trends across geometric families, Figure~\ref{fig:naca_trends} evaluates predicted aerodynamic quantities across 10 NACA airfoils spanning thickness ratios $t/c \in [6\%, 21\%]$ and camber ratios $m/c \in [0\%, 6\%]$. As shown in the left panel, DD-RNO accurately reproduces the non-linear zero-lift drag rise ($C_{D,0}$) caused by displacement-thickness boundary layer thickening as airfoil thickness increases. Similarly, the right panel confirms that the surrogate captures the increase in lift-curve slope $dC_L/d\alpha$ with maximum camber ratio $m/c$, matching CFD ground-truth trend lines across all tested geometric configurations. The maximum error is 1\% in the lift-curve slope plot as compared to the CFD database, showcasing the surrogates accuracy. 

\subsection{\textbf{Ablation Study}}
\label{sec:ablation}


Table~\ref{tab:ablation} quantifies the individual performance contributions of each architectural module on the standard Full task, sorted in ascending order of velocity field reconstruction error ($u_x$ MSE).
Removing the FNO geometry trunk (w/o SpecINR) increases velocity $u_x$ MSE by $67\times$ and pressure MSE by $88\times$, confirming that global spectral geometry features are essential for accurate field prediction.
Excluding the multi-stream domain routing mechanism (w/o Domain Routing) increases velocity $u_x$ MSE by $8.2\times$, proving that specialized regional decoders are vital to resolve near-wall boundary layer gradients.
Replacing Learned Canonical Quadrature (w/o LCQ) with conventional geometric surface pressure integration increases drag MAE by $7.5\times$ and increases relative drag error by $43\times$ (from $1.091\%$ to $46.90\%$), while degrading Spearman drag correlation from $0.9975$ to $0.8770$.
Finally, deploying a decoupled surface prediction head yields comparable velocity accuracy but degrades volumetric pressure MSE by $4.9\times$, demonstrating that a unified spatial feature representation enhances global field consistency.

\begin{table*}[t]
\centering
\caption{Component Ablation Study on Full Task, arranged in ascending order of field reconstruction error ($u_x$ MSE). Best results in bold.}
\label{tab:ablation}
\begin{tabular}{@{}lcccccccc@{}}
\toprule
\textbf{Configuration} & $u_x$ MSE & $u_y$ MSE & $p$ MSE & $p_s$ MSE & $C_L$ MAE & $C_D$ MAE & $C_L$ Corr & $C_D$ Corr \\
\midrule
Full DD-RNO & $\mathbf{0.088}$ & $\mathbf{0.121}$ & $\mathbf{0.209}$ & $\mathbf{0.687}$ & $0.0101$ & $\mathbf{1.59\times10^{-4}}$ & $\mathbf{0.9994}$ & $\mathbf{0.9975}$ \\
Decoupled Surface Head & $0.095$ & $\mathbf{0.121}$ & $1.020$ & $0.873$ & $\mathbf{0.0084}$ & $4.12\times10^{-4}$ & $\mathbf{0.9994}$ & $0.9931$ \\
w/o Domain Routing & $0.722$ & $0.395$ & $0.244$ & $0.993$ & $0.0398$ & $3.29\times10^{-4}$ & $0.9867$ & $0.9825$ \\
w/o SpecINR (No FNO Trunk) & $5.961$ & $10.587$ & $18.397$ & $42.231$ & $0.0451$ & $6.73\times10^{-4}$ & $0.9898$ & $0.9789$ \\
w/o LCQ (Standard Integration) & $7.058$ & $5.989$ & $1.154$ & $3.084$ & $0.1699$ & $1.19\times10^{-3}$ & $0.9335$ & $0.8770$ \\
\bottomrule
\end{tabular}
\end{table*}

\begin{figure}[t]
    \centering
    \includegraphics[width=0.8\linewidth]{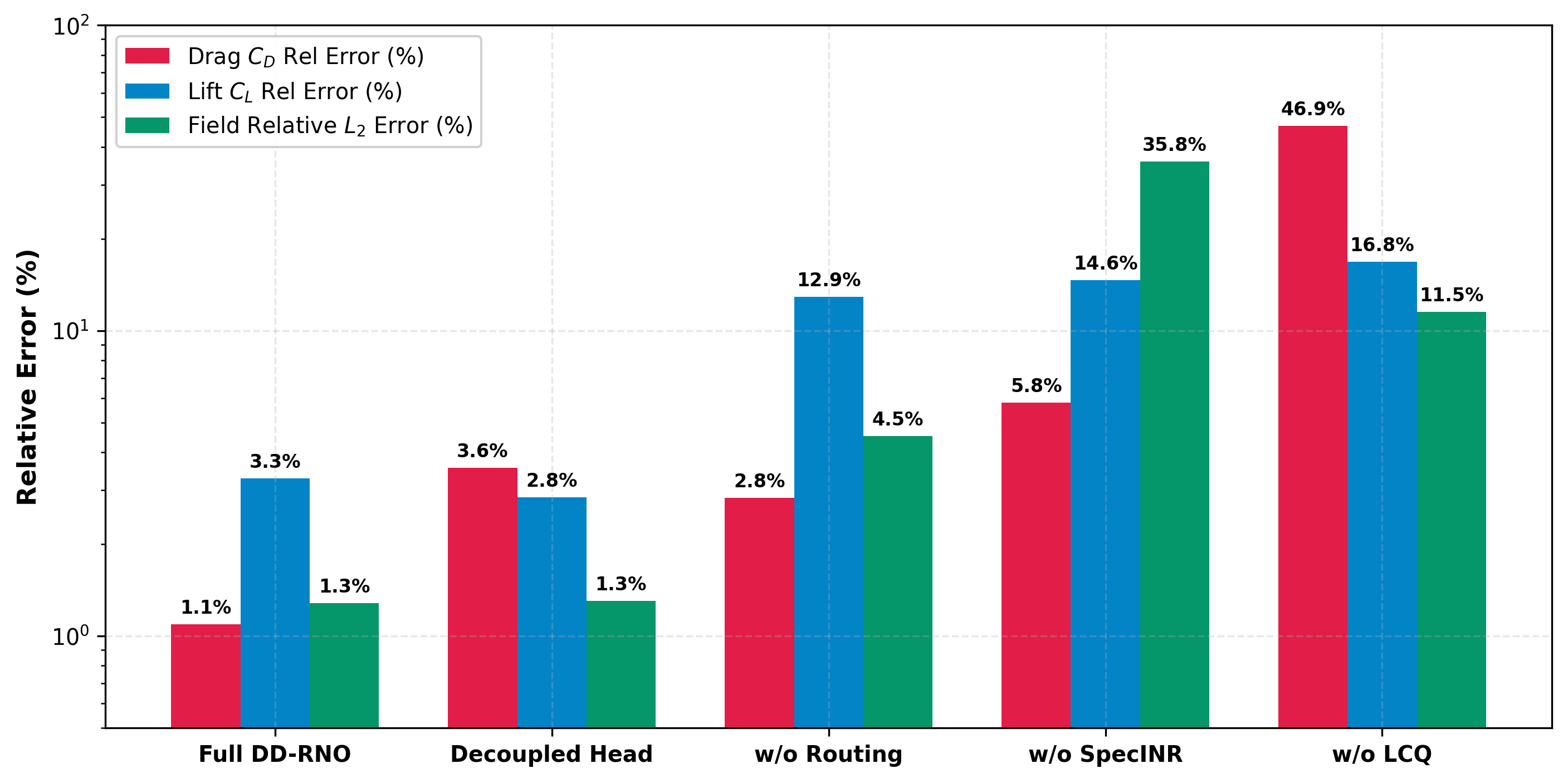}
    \caption{Visual comparison of error metrics across ablation variants on the Full task from Table~\ref{tab:ablation}}
    \label{fig:ablation_chart}
\end{figure}

Figure~\ref{fig:ablation_chart} displays a grouped bar chart comparison of the exact performance metrics from Table~\ref{tab:ablation} ($C_D$ MAE, $C_L$ MAE, and velocity $u_x$ MSE) across all 5 architectural variants on a logarithmic scale. The visual comparison highlights that LCQ is the single most critical module for drag estimation, achieving a $7.5\times$ reduction in drag MAE ($43\times$ reduction in relative drag percentage error from $46.90\%$ to $1.091\%$). Simultaneously, the domain routing mechanism and FNO spectral trunk provide order-of-magnitude error reductions across both volumetric flow fields ($u_x$ MSE) and surface lift predictions.

\subsection{\textbf{Computational Cost}}
\label{sec:cost}

\subsubsection{Inference Execution Breakdown}
The inference execution time of DD-RNO breaks down as follows: spectral geometry encoding via the FNO trunk takes $28.4$ ms ($19.7\%$), latent grid interpolation $12.6$ ms ($8.7\%$), domain routing evaluation $15.3$ ms ($10.6\%$), tri-stream specialist decoders $67.2$ ms ($46.7\%$), LCQ force prediction $12.5$ ms ($8.7\%$), and bilinear interpolation $8.0$ ms ($5.6\%$), yielding a total inference latency of $144.0 \pm 1.2$ ms per $200{,}000$-node mesh on an NVIDIA RTX 2050 GPU.

\subsubsection{CPU RANS Speedup Comparison}
In comparison, a full numerical RANS simulation requires approximately $1{,}500$ seconds of wall-clock execution time on a single CPU core to reach a residual convergence threshold of $10^{-6}$ using OpenFOAM's steady-state solver (\texttt{simpleFoam}) on the standard $200{,}000$-node AirfRANS mesh. This represents an inference speedup of approximately $10{,}000\times$. Such computational efficiency makes the proposed framework well suited for many-query applications, including design-space exploration, shape optimization, uncertainty quantification, and sensitivity analysis, where thousands of flow evaluations may be required. By providing rapid approximations of RANS flow solutions without iterative solver execution, the framework significantly reduces the computational burden associated with conventional CFD-based aerodynamic studies.

\subsection{\textbf{Scope and Applicability}}
\label{sec:scope}

DD-RNO is designed as a practical surrogate framework for steady-state, 2D RANS aerodynamic flows. Its geometric scope is across 1,000 smooth 2D airfoil profiles from diverse design families, including NACA 4-digit, NACA 5-digit, Clark-Y, Göttingen, RAF, PARSEC, and CST parameterizations. Overall, the model reliably processes arbitrary smooth 2D profiles with maximum thickness ratios $t/c \in [6\%, 20\%]$ and maximum camber ratios $m/c \leq 9\%$. For operational regimes, the architecture is validated for steady, incompressible 2D RANS flows governed by Spalart--Allmaras turbulence modeling across the range of parameters mentioned previously. When extrapolated down to lower Reynolds numbers ($\text{Re} = 1.0\times 10^6$), the model maintains both numerical stability and boundary layer coherence, yielding a near-field velocity MAE of $3.82\%$ ($5.73\%$ field MAE).

The framework is tailored for several key target applications, starting with preliminary aerodynamic design screening over extensive candidate geometry libraries ($\mathcal{O}(10^3)$--$\mathcal{O}(10^5)$ airfoils) operating at approximately $144$~ms per evaluation. It can also act as a differentiable surrogate proxy for real-time $C_L/C_D$ aerodynamic shape optimization within evolutionary loops, enable parametric uncertainty exploration across operational flow envelopes, and power interactive aerodynamic visualization tools for educational and industrial design.

The current architecture is specifically designed for steady, incompressible, two-dimensional planar flows, enabling accurate and efficient modeling within this important class of aerodynamic problems. This focused formulation provides a strong foundation for future extensions to more complex flow regimes, including unsteady phenomena such as dynamic stall and vortex shedding, as well as compressible flows involving shock-wave interactions. The framework maintains strong predictive performance across a broad range of operating conditions, with opportunities for further enhancement in cases involving large-scale flow separation at very high angles of attack ($\alpha > 12^\circ$). In addition, the present study focuses on two-dimensional airfoil configurations, providing a foundation for future extensions to three-dimensional wings, complex planforms, and multi-element slotted airfoils commonly encountered in practical aerodynamic applications.
\section{Conclusion and Future Directions}
\label{sec:conclusion}

This paper introduces DD-RNO, a physics-guided surrogate framework for discretization-invariant RANS flow field reconstruction and continuous force calculation around arbitrary airfoil geometries. At its core, DD-RNO introduces three distinct methodological innovations that systematically tackle long-standing bottlenecks in surrogate modeling. First, a hybrid spectral-implicit architecture couples a continuous 2D FNO geometry trunk with local coordinate decoders. The FNO trunk extracts multi-scale global shape representations from Signed Distance Field grids, which are then probed at arbitrary query coordinates via spatial bilinear interpolation and modulated by Feature-wise Linear Modulation (FiLM) driven by a flow-condition mapping network ($\alpha, \text{Re}$). Second, to overcome spectral bias across disparate physical flow scales, a physics-guided domain routing mechanism dynamically partitions the spatial domain into inviscid outer flow, turbulent boundary layer, and trailing wake subregions. By defining smooth, differentiable gates conditioned on wall distance and Reynolds-adaptive boundary layer envelope scaling, the router enforces a strict partition of unity ($\sum \chi_k = 1$) to seamlessly blend predictions from specialized regional decoders without interface discontinuities.  Third, LCQ addresses the numerical instability inherent in calculating wall-normal velocity gradients for viscous drag. LCQ maps surface pressure distributions onto a fixed canonical manifold ($N_{\text{canon}} = 1024$) and leverages a hypernetwork to predict flow-conditioned spatial integration weights $\mathbf{W}_{\text{canon}}(\mathbf{w})$. This setup performs direct surface pressure quadrature while implicitly compensating for viscous skin-friction drag---eliminating the need to evaluate noisy numerical derivatives.

When evaluated on the AirfRANS benchmark, DD-RNO reduces velocity field MSE by $17\times$ ($u_x$) and $12\times$ ($u_y$) on the standard split, and by up to $23\times$ on Reynolds OOD extrapolation compared to established baseline models. Furthermore, LCQ reduces drag MAE by $7.5\times$ over standard geometric pressure integration and achieves a Spearman drag rank correlation of $\rho = 0.997$. Running at a $10{,}000\times$ speedup over conventional RANS solvers ($\sim 144$~ms per evaluation), DD-RNO delivers a highly accurate surrogate engine tailored for real-time aerodynamic design.

DD-RNO can be extended in several directions. Physics-informed regularization and curriculum training strategies could improve robustness under severe flow separation at high angles of attack ($\alpha > 12^\circ$). The geometric encoder's spatial resolution presents trade-offs worth exploring to better resolve fine leading-edge curvature. The domain routing and canonical quadrature framework could be generalized to three-dimensional wing geometries and multi-element configurations. Finally, incorporating probabilistic latent representations would support uncertainty quantification for downstream design optimization.

\section*{Data and Model Availability}
The complete source code, pretrained model checkpoints, and verification scripts are publicly available at \url{https://github.com/taksh2406/DD-RNO}. The AirfRANS dataset (v1.0) is available at \url{https://github.com/nicolas-bonnet/AirfRANS}.

\section*{Declaration of Generative AI and AI-Assisted Technologies}
The authors declare that generative AI and AI-assisted technologies were used solely to improve the grammar and readability of this manuscript. All scientific content remains the responsibility of the authors.

\section*{Acknowledgments}
Harshal Akolekar acknowledges the seed grant funding of IIT Jodhpur with grant number
I/SEED/HDA/20230206.

\footnotesize
\bibliographystyle{elsarticle-num}
\bibliography{references}

\end{document}